\documentclass[a4paper,11pt]{article}
\usepackage{jinstpub} 
\usepackage{lineno}
\usepackage{textcomp}
\usepackage{subcaption}
\usepackage{url}
\usepackage{hyperref}

\title{Mitigating Sensor Cracking using the Interposer Strategy in the ATLAS ITk Strip Detector}

\collaboration{The ATLAS ITk Strips Collaboration}

\emailAdd{anne.winifred.fortman@cern.ch}

\abstract{This paper introduces the strategy of using an interposing layer to mitigate thermal-stress-induced fracturing of silicon sensors in ATLAS ITk Strip modules. This strategy involves a change in the module design in which layers of soft adhesive and a kapton separator, known as an interposer, are added to the module stack-up to decouple thermal stresses. This paper discusses the interposer design and its implementation in the ATLAS ITk Strip barrel, including material tests to validate the additional components, the development of new processes to integrate the new layers into the module assembly workflow, results from design validation and quality control testing of prototype interposed modules, and studies indicating that the new design significantly reduces the thermal stress experienced by the sensor. Tests of prototype support structures loaded with interposed modules have shown no sensor fractures in the relevant temperature range for operation in ATLAS, resulting in the adoption of the interposer strategy as the primary sensor cracking mitigation in the ATLAS ITk Strip barrel.}


\clearpage

\begin{document}
\maketitle
\flushbottom

\section{Introduction to the Sensor Cracking Problem}
\label{sec:intro}

\subsection{The ATLAS ITk Strip Detector}

\begin{figure}[htbp]
\centering
\includegraphics[width=.7\linewidth]{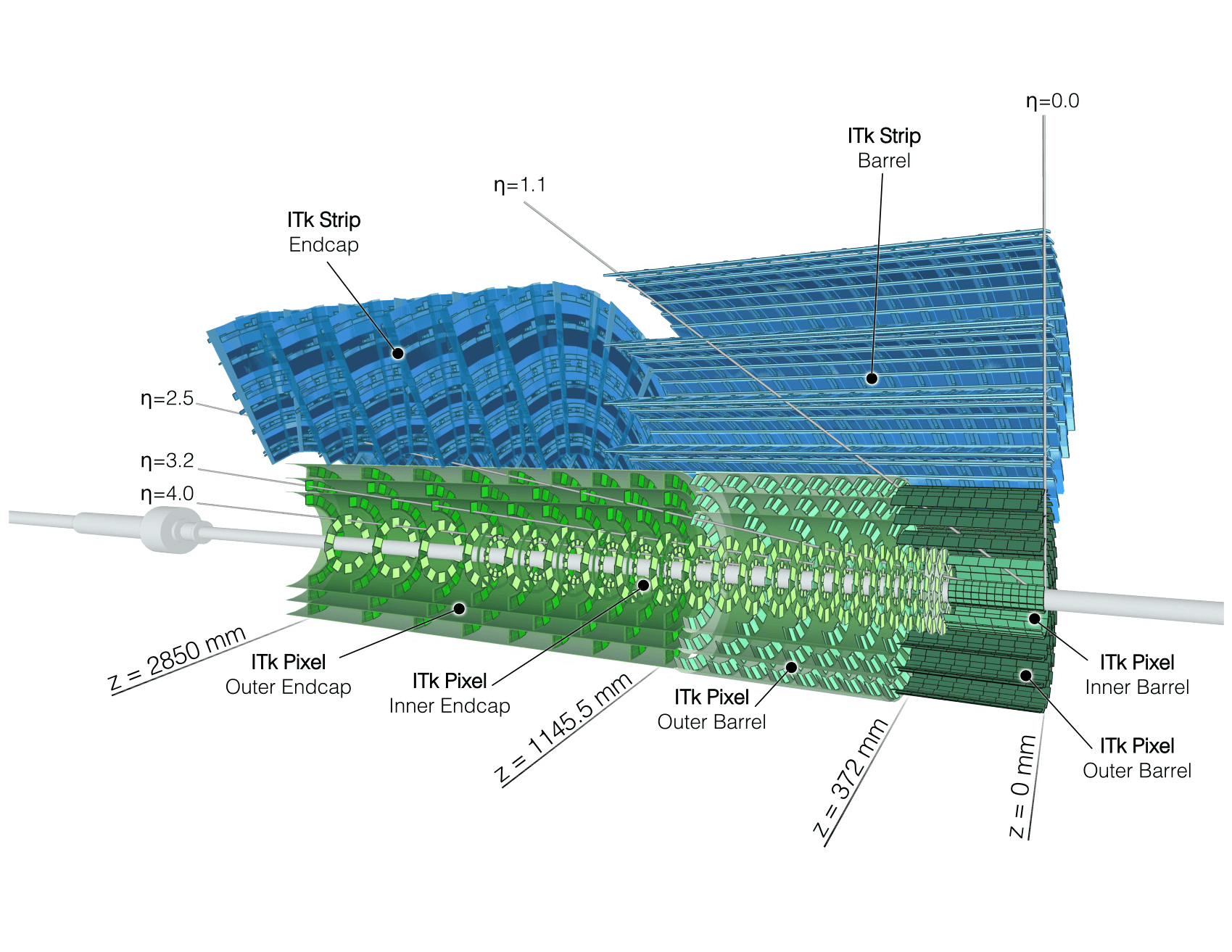}
\caption{A diagram of one section of the ATLAS ITk detector. The inner layers make up the ITk Pixel detector. The outer layers form the ITk Strips system. ATLAS uses a right-handed coordinate system with its origin at the nominal interaction point (IP) in the center of the detector and the \(z\)-axis along the beam pipe. The \(x\)-axis points from the IP to the center of the LHC ring, and the \(y\)-axis points upwards. Polar coordinates \((r,\phi)\) are used in the transverse plane, \(\phi\) being the azimuthal angle around the \(z\)-axis. The pseudorapidity is defined in terms of the polar angle \(\theta\) as \(\eta = -\ln\tan(\theta/2)\).  \cite{ITkFig}}
\label{fig:ITk}
\end{figure}

The Inner Tracker (ITk), depicted in Figure~\ref{fig:ITk}, is an all-silicon tracker comprising two tracking subsystems and is an upgrade to the ATLAS detector \cite{ATLASref} for the High-Luminosity LHC \cite{Aberle:2749422}. The ITk Pixel detector \cite{PixelTDR} will make up the inner layers of the ITk, while the ITk Strip detector \cite{StripTDR} forms the outer layers. The ITk Strip tracker is laid out in a cylindrical structure with 4 double-sided concentric layers covering the barrel region bookended by 6 ring-shaped layers on each side covering the endcap region. 

Although the sensor cracking problem and the mitigation strategy described in this paper are common to both the barrel and the endcap regions, there are significant differences in the implementation of the strategy in the two regions due to geometrical differences. This paper describes the implementation in the barrel, and a future paper will detail the implementation in the endcap region.

\begin{figure}[htbp]
\centering
\includegraphics[width=.99\textwidth]{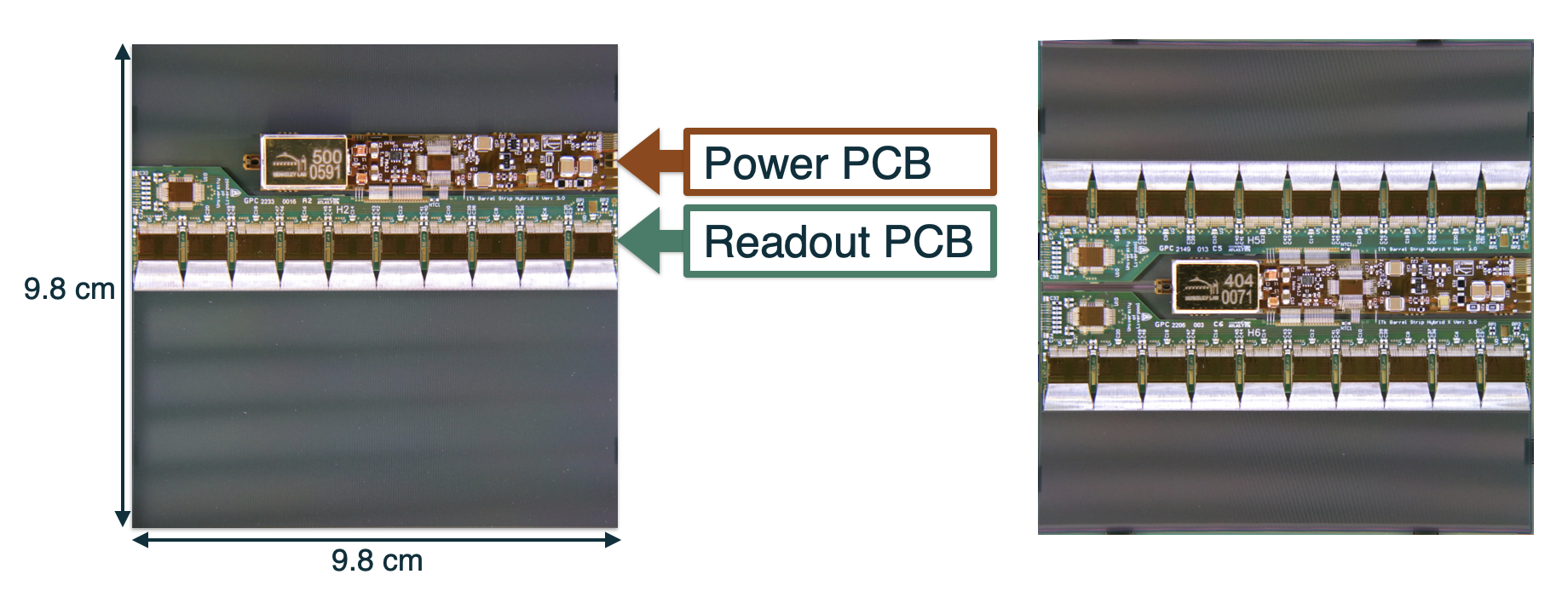}
\caption{Photographs of a long-strip module (left) and a short-strip module (right).\label{fig:modules}}
\end{figure}

\begin{figure}[htbp]
\centering
\includegraphics[width=.5\textwidth]{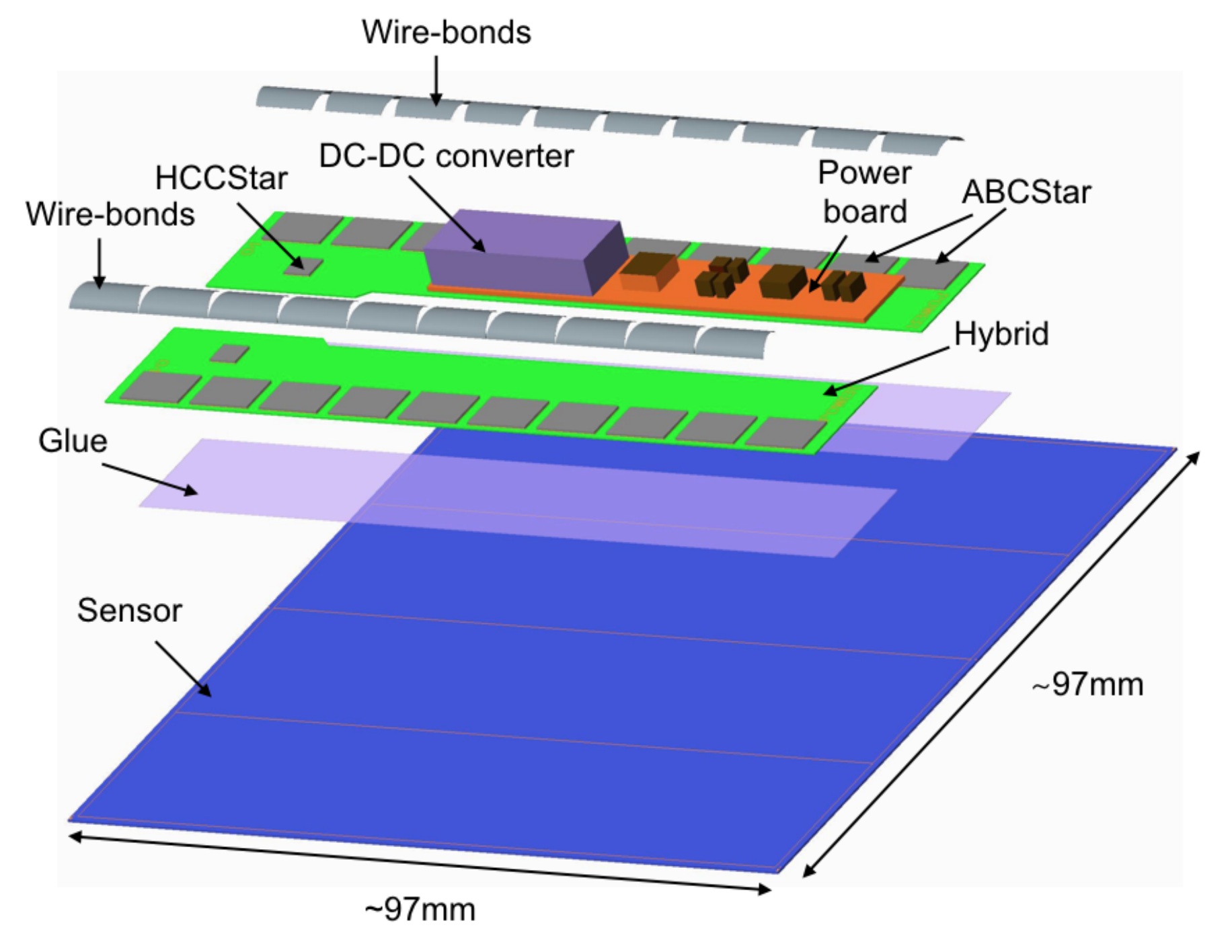}
\caption{An exploded view of the components of a short-strip module \cite{StripTDR}}
\label{fig:explodedmodule}
\end{figure}

Each layer of the ITk Strip detector will be tiled with \textit{modules}. A module is composed of front-end electronics glued to a silicon microstrip sensor. The ITk Strip barrel will contain a total of 10,976 modules of two varieties: The inner two barrel layers will contain \textit{short-strip} modules, while the outer two layers will be composed of \textit{long-strip} modules \cite{ABC130paper, StripTDR}. A photograph of each module type is shown in Figure~\ref{fig:modules}. An exploded view of a short-strip module is shown in Figure~\ref{fig:explodedmodule}. 

Each sensor is made of n+-doped strip implants in p-type silicon bulk, and measures $\sim$9.8 cm by $\sim$9.8 cm \cite{StripTDR, ABC130paper}. Long-strip sensors contain 2 rows of $\sim$5-cm-long strips, totaling 2,560 strips per sensor. Short-strip sensors contain double the amount of strips, with 4 rows of $\sim$2.5-cm-long strips. The strip pitch for each sensor type is $75.5$ $\mu$m.

Front-end electronics are mounted on flexible printed circuit boards (\textit{flex PCBs}). The front-end electronics on each readout PCB consist of the HCCStar and ABCStar readout ASICs \cite{ABC130chip} that process and read out data. Long-strip modules require one readout PCB and short-strip modules require two readout PCBs to read out data from all strips. Additionally, a power PCB provides power to the readout ASICs on each module. Each readout PCB and power PCB is fixed to the sensor with a rigid epoxy (Loctite Eccobond F 112 \cite{TrueBlue}). Electrical connections between the front-end electronics and the sensor, and between the ASICs and the flex PCBs, are made with wirebonds.

\begin{figure}[htbp]
\centering
\includegraphics[width=\textwidth]{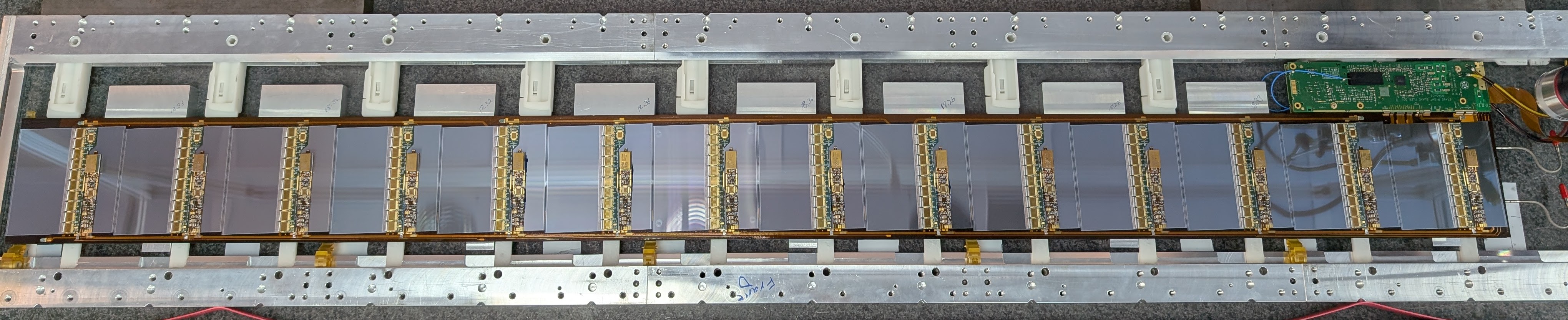}
\caption{A stave loaded with long-strip modules. $14$ modules are glued to each side of the stave.}
\label{fig:stave}
\end{figure}

Modules are glued to rigid carbon-fiber support structures known as \textit{staves}. Staves provide mechanical stability, distribute cooling, and carry power and data readout lines to each module. A photo of a stave loaded with modules is shown in Figure~\ref{fig:stave}. Further details concerning staves can be found in \cite{SergioStavepaper}.

Overall, the ITk Strip detector was designed as a simple modular system with a minimum material budget. Gluing the front-end electronics directly to the sensor rather than including an extra material part above or to the side of the sensor allows near-complete coverage of each stave side with active detector material and a reduction in detector mass. This configuration also allows cooling of front-end electronics through thermal contact with the sensor, which is glued to the carbon-fiber stave containing an embedded evaporative CO$_2$ cooling system \cite{ARLING2022166953}. The ITk Strip detector is planned to operate at temperatures down to $-35$\textdegree C to maintain charge collection efficiency of the sensors in the harsh radiation environment. Repeated warm-ups to approximately room temperature are expected during year-end technical stops.

\subsection{Sensor Cracking due to Thermal Stress}

During the ramp-up to module production, modules were attached to staves and stress-tested by thermal cycling between $20$\textdegree C and $-35$\textdegree C. After thermal cycling, cracks were observed in some sensors.

\begin{figure}[htbp]
\centering
\includegraphics[width=\textwidth]{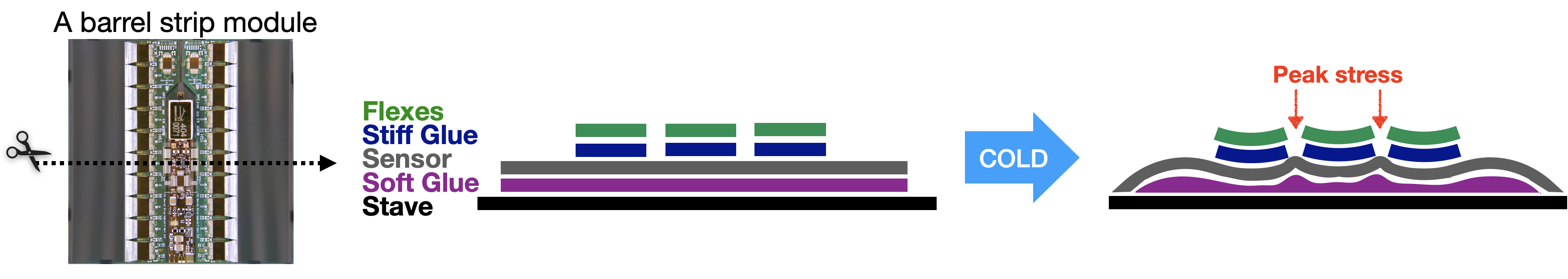}
\caption{A cartoon of a barrel short-strip module cross-section is shown at room temperature and while cold. At cold temperatures, the mismatch between the coefficients of thermal expansion of the flex PCBs and the sensor introduces thermal forces that can lead to cracks.}
\label{fig:crackingsketch}
\end{figure}

Figure~\ref{fig:crackingsketch} demonstrates the mechanism that induces stress on the sensor. The coefficient of thermal expansion (CTE) of the readout and power PCBs, primarily driven by layers of copper in the flex PCBs, differs from the CTE of the silicon sensor. Assembly of modules and staves occurs at room temperature. When modules are cooled, the flex PCBs contract more than the silicon, exerting force on the sensor through the stiff epoxy layer between them. The silicon sensor is tacked to the rigid carbon-fiber stave using a soft adhesive (Dow SE 4445 \cite{SE4445}) that does not prevent the sensor from deforming. As a result, the loaded sensor experiences stress and deformation from the contracting flex PCBs that can lead to cracks.

Assessing potential CTE mismatch is standard practice in detector design. During the module design and prototyping phase, module thermal stress was evaluated using a finite-element thermo-mechanical model. These early simulations predicted peak sensor stresses within acceptable limits. Since then, several adjustments have been made to the module design, including a change in the particular epoxy used and in the epoxy glue patterns between the flex PCBs and the sensor. A more detailed finite-element thermo-mechanical model was developed after the cracking problem was discovered \cite{Vallone:2938828}. This model reveals that the details of the epoxy glue pattern can give rise to highly localized stress peaks, demonstrating that these design changes may have had significant impact on the thermal stress.

In addition, failures through mechanical stress induced by CTE mismatch were assumed to be detectable during the routinely-performed thermal cycling of modules prior to loading on staves. At the time, it was not understood that modules loaded onto staves would actually experience higher mechanical stress than modules not glued to staves. Fractures were not observed in thermal-cycled modules prior to loading on staves. Furthermore, stress-induced fractures in early modules loaded on staves presented similar symptoms as common electrical failures or isolated problems due to mishandling, and therefore were initially misidentified. Only with increasing statistics of early modules loaded to staves and operated under extreme temperature conditions did thermal stress-induced cracking become evident as a systemic problem, delaying the detection of the issue.

\begin{figure}[htbp]
\centering
\includegraphics[width=0.5\textwidth]{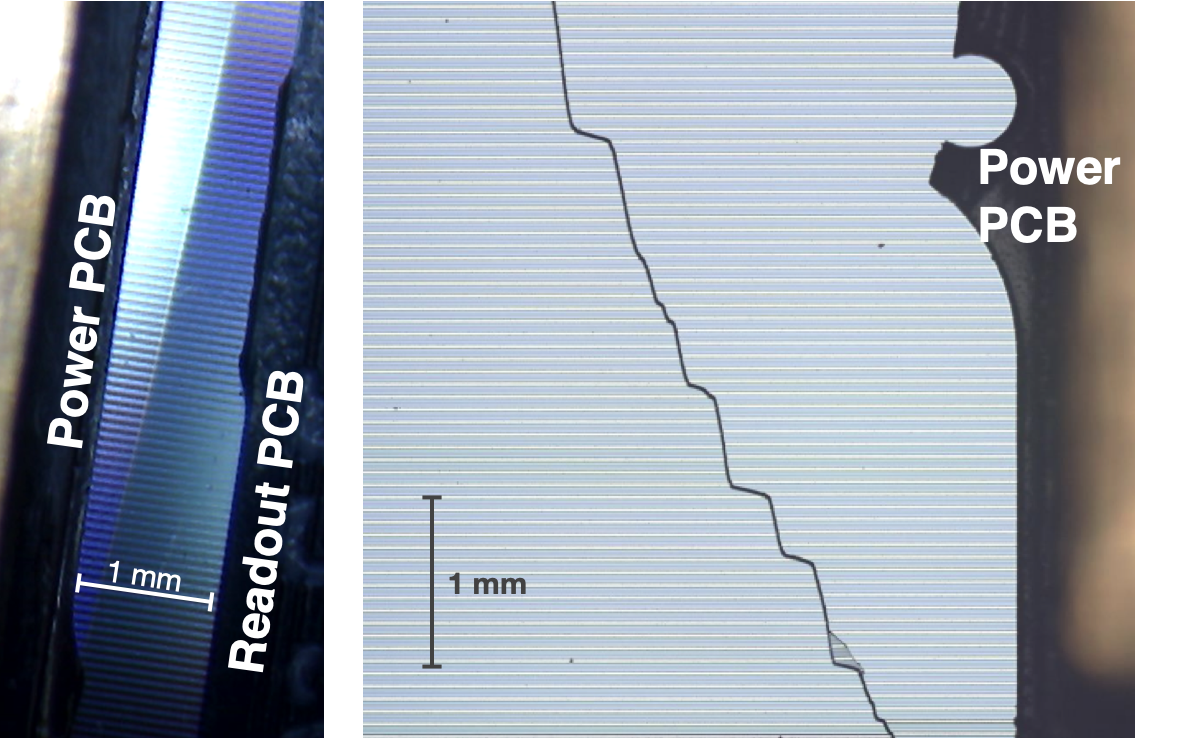}
\caption{Photos of cracks in sensors that were mounted to staves and brought to cold temperatures, inducing thermal stress on the sensor. Left: a crack is visible on the sensor in the 1-mm gap between the power PCB and the readout PCB. Right: on a different sensor, a crack in the sensor surface is visible near the power PCB.}
\label{fig:crackphotos}
\end{figure}

Figure~\ref{fig:crackphotos} shows examples of cracks in sensors. Cracks often appear in the $\sim$1-mm gap between the flex PCBs, in the vicinity of the flex PCBs, or at the edges of the epoxy adhesive. The cracks can be very small, and in some cases partially or fully hidden beneath wirebonds or flex PCBs, making them difficult to spot during visual inspection. Instead, cracks are typically identified through anomalous noise behavior or sensor currents during module electrical testing \cite{AbeCrackingProc}.

A crack in the sensor creates a low-resistivity path between electrodes, resulting in increased leakage current or current instability that ultimately prevents full and stable depletion of the sensor in the detector environment. In the few cases where partial depletion is still possible, sensors with stress-induced fractures have not been found to be reliable in longer-term operation. For this reason, even a very small crack through only a few strips means that the entire module should be switched off during detector operation, reducing tracking efficiency.

The lowest planned operating temperature for an ITk Strip module is $-35$\textdegree C. However, during their expected 10-year lifespan, modules may experience lower temperatures due to power or cooling failures. The lowest temperature that a module could be exposed to would be roughly $-55$\textdegree C in the case of unplanned venting of the evaporative CO$_2$ cooling system \cite{StripTDR}. The relevant temperature range that modules should ideally endure without cracking is therefore [$20$\textdegree C$,-55$\textdegree C].

If no action were taken to mitigate the sensor cracking problem, it is estimated from statistics of early modules loaded to staves that roughly $20$\% (sample size of $42$ modules) of barrel modules would crack at the nominal operating temperature, and up to roughly $70$\% (sample size of $14$) of the barrel would crack in the case of a burst cooling pipe. The start of module production was therefore postponed until a suitable cracking mitigation strategy could be found. 

This paper introduces the interposer mitigation strategy that will be used in both the ITk Strip barrel and endcap, and its implementation in the ITk Strip barrel. First, Section~\ref{sec:solutionsection} discusses possible mitigation strategies and motivates the redesign of modules according to the interposer strategy. Next, Section~\ref{sec:designsection} unveils the new module design and describes the material testing studies that shaped and validated it. Section~\ref{sec:arrayinterposing} addresses the practical challenges of building redesigned modules and integrating the interposer strategy into the established module production workflow. Before committing to the interposer strategy, it was necessary to show that prototype redesigned modules pass design validation and quality control criteria, as demonstrated in Section~\ref{sec:prototypesection}. Section~\ref{sec:stresssection} presents evidence of sensor stress reduction in prototype redesigned modules, supporting the adoption of the interposer strategy in the detector. Section~\ref{sec:stavesection} summarizes the promising preliminary results of redesigned modules mounted on staves. Finally, Section~\ref{sec:outcomesection} discusses the outlook for the ITk Strip barrel with the interposer strategy in place. This paper extends the original module design proposed in the Technical Design Report \cite{StripTDR}.

\section{The Interposer Strategy}
\label{sec:solutionsection}

By the time the sensor cracking issue was discovered, the ITk Strip project was transitioning to full module production. Any potential cracking mitigation strategy would need to cause minimal interruption to the existing module production workflow. Any extra materials incorporated would need to have been previously qualified for use in the ITk environment according to procedures similar to those described in \cite{Poley_2016}. Significant changes to the design of any components would introduce delay, as years of testing had been required to validate existing designs. Several potential sensor stress mitigation strategies were explored within these constraints, including changes to the module loading glue and to the module design.

\begin{figure}[htbp]
\centering
\includegraphics[width=\textwidth]{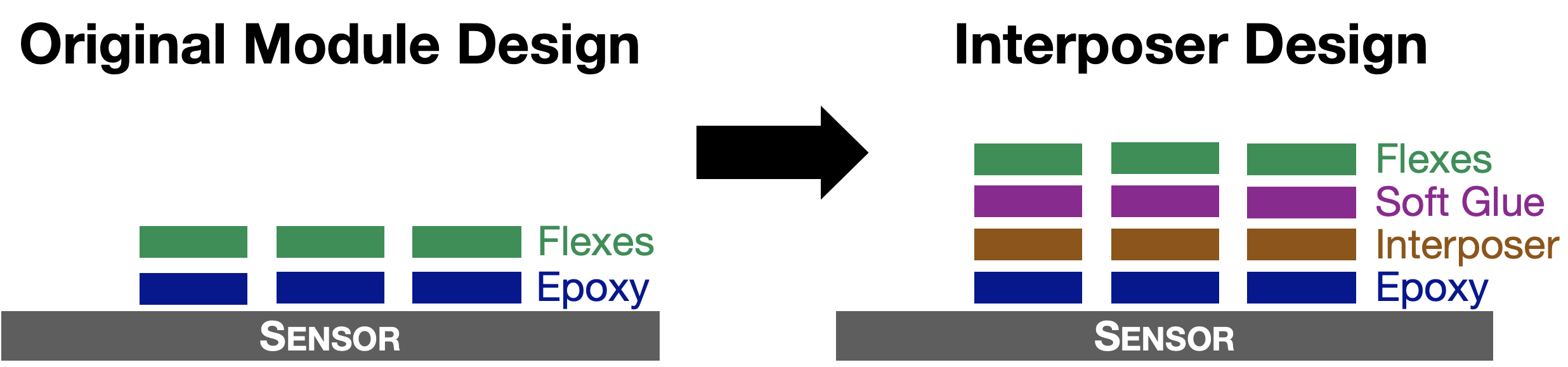}
\caption{A comparison of the original module design with the interposer design. The interposer design adds two new layers beneath each flex PCB: a layer of soft glue and a layer of material referred to as an interposer.}
\label{fig:interposerconcept}
\end{figure}

The strategy introduced in this paper is to include extra layers of material between each flex and the sensor, as shown in Figure~\ref{fig:interposerconcept}. The main stress-relieving mechanism is provided by the new layer of soft glue beneath each individual flex PCB: the soft glue absorbs stress from the thermal contraction of the flex, decoupling the thermal contraction of the flex PCBs from the sensor.

The adhesive already in use to attach modules to staves was an obvious candidate soft glue, as it had been previously studied and found to meet the radiation qualifications for use in ITk. Dow SE 4445 \cite{SE4445}, a two-part thermally conductive gel\footnote{In this paper, the term glue is used synonymously with adhesive to mean "a material that binds other materials together by surface attachment" \cite{SHIELDS1984336}. A gel is a semi-solid system that cures to a soft consistency. Silicone-based gels are commonly used in electronics applications with high temperature variation to protect components against thermal and mechanical stress \cite{siliconegels}. In this application, the SE 4445 gel is used as an adhesive and will be referred to as a glue for the purposes of this paper.}, is significantly softer than the rigid epoxy originally used between the flex PCBs and the sensor. Commercial measurements of SE 4445 were made indicating a storage modulus of 0.5 MPa, while the modulus of a typical epoxy is 3-5 GPa. 

SE 4445 was also known to alleviate a second issue facing ITk Strip modules, known as \textit{cold noise} (described further in Section~\ref{sec:QA}). During past studies, cold noise had been shown to significantly improve when replacing the original epoxy with SE 4445 in the module stack-up. Using this glue in the new design was therefore seen as an opportunity to simultaneously address cold noise. However, these studies also found that direct contact between SE 4445 and the sensor surface leads to unstable high bias currents.

For this reason, the SE 4445 layer in Figure~\ref{fig:interposerconcept} does not replace the original epoxy in contact with the sensor, but is instead placed above the interposer layer. The epoxy and the interposer layer separate the SE 4445 from the sensor surface. The interposer layer is further needed to separate the two glues since it is known \cite{SVENDSEN2007678} and was observed that the epoxy fails to cure or bond properly in direct contact with the silicone-based gel.

Variations on the interposer configuration shown in Figure~\ref{fig:interposerconcept} were also considered. Explicitly requiring a rigid interposer layer (silicon, for example) could provide an alternative stress-relieving mechanism: the rigid material could stiffen the module and act as a shield, diverting stress away from the sensor. The soft glue layer could optionally be replaced with a second layer of the original epoxy above a rigid interposer, simplifying the design by not introducing a new adhesive.

A finite element thermo-mechanical model was developed to simulate the assembly, cool-down, and operation of the modules \cite{Vallone:2938828, simulation}. This simulation was used to evaluate predicted peak sensor stress under various potential changes to module design or module loading, informing the overall cracking mitigation strategy. Various configurations of module stack-ups were evaluated in the simulation.

According to the simulation, the stress-decoupling mechanism offered by the soft glue layer dominates the sensor stress reduction provided by any interposer configuration, outweighing any stress-shielding effect of a rigid interposer. Regardless of the particular interposer material modeled, for a thin interposer (< 300 $\mu$m thick), the design with SE 4445 above the interposer layer was predicted to reduce peak sensor stress by 90-95\%, virtually eliminating the concern of thermal-stress-induced cracks \cite{Vallone:2938828, simulation}. The interposer configuration shown in Figure~\ref{fig:interposerconcept} was therefore preferred over other variations.

Although this predicted stress reduction was the most promising of any strategy that had been modeled, the interposer strategy required changing the module design. Implementing this strategy would require extensive material testing to validate the design, new tooling and assembly methods to integrate the interposer layers, and a program of design validation and quality control testing of prototype modules. Therefore, the interposer strategy was initially developed as a backup option while simpler mitigation strategies were tested first. 

For example, the SE 4445 attaching the sensor to the stave could be replaced with a rigid glue (Henkel Loctite EA 9396 AERO Epoxy \cite{Hysol}), improving sensor support from below and requiring no change to the module design. This strategy was tested with several staves that were loaded using the rigid glue and thermal cycled. A significant sensor cracking rate was observed while thermal cycling to relevant temperatures, and this strategy was not pursued further \cite{AbeCrackingProc, damen2025staveresultsmitigatingsensor}.

As other stress mitigation strategies proved insufficient, the interposer design became the most feasible remaining strategy and was developed in earnest.

\section{Design and Validation of the Interposer Strategy}
\label{sec:designsection}

\subsection{The Standard Interposed Module Design}

As the simulation predicted that the stress decoupling effect was independent of the particular interposer material, the project was free to consider the most practically convenient options for the interposer layer. During material testing and development of the interposer design, non-adhesive kapton HN \cite{KaptonHN} was considered the preferred material. Kapton is cheap and widely available, and could be easily and cheaply cut to match the shapes of the flex PCBs. 
 
Another advantage of kapton is its availability in <$100$ $\mu$m thicknesses, reducing the material added to the module. Increasing the overall module height presented a concern, as only several hundred microns of clearance were available to avoid collisions with ITk infrastructure during installation. As a result, kapton thicknesses of 25 and 50 $\mu$m were initially considered. However, due to difficulty handling 25 micron kapton, 50-$\mu$m-thick kapton was ultimately chosen. A target thickness of 100 $\mu$m for the SE 4445 layer was similarly chosen to minimize added material to the module while retaining its stress-relieving effect according to simulation.

\begin{figure}[htbp]
\centering
\includegraphics[width=0.4\textwidth]{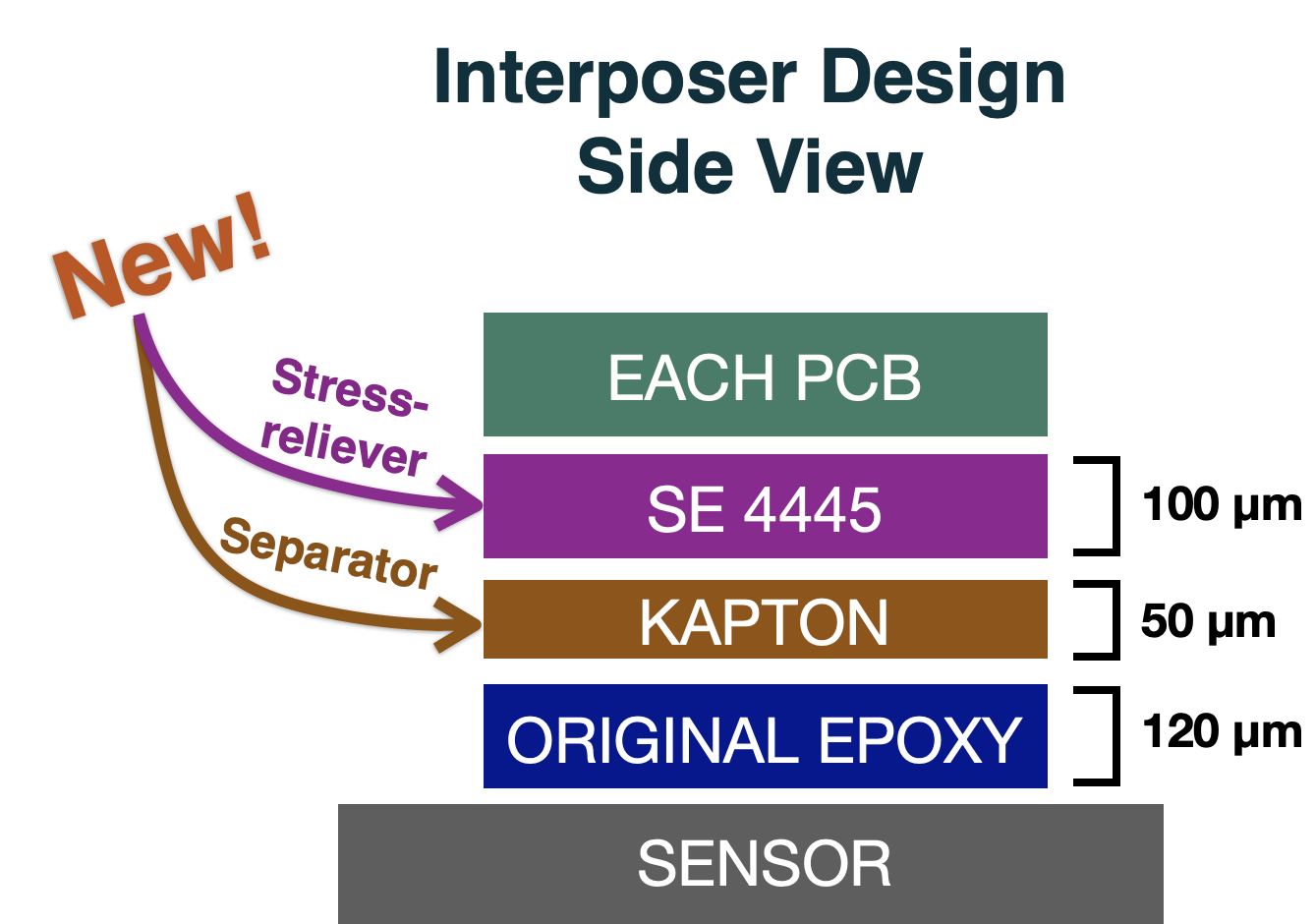}
\caption{The standard interposer design. A layer of soft glue and a layer of kapton underneath each individual flex PCB are added to the original module design \cite{AnneInterposingProc}.}
\label{fig:interposerdesign}
\end{figure}

The standard interposed module design is shown in Figure~\ref{fig:interposerdesign}. Between each flex PCB and the sensor, the new module design is composed of:

\begin{itemize}
    \item A 100-$\mu$m-thick layer of SE 4445
    \item A 50-$\mu$m-thick layer of kapton 
    \item The original 120 $\mu$m layer of epoxy (unchanged).
\end{itemize}

The additional material budget of ITk strip modules due to the standard interposer design was estimated and found to be negligible. The 50-$\mu$m kapton layer under the flex PCBs adds an estimated 0.005\% of a radiation length to the material budget of a short-strip module, whereas the SE 4445 glue adds an estimated 0.25\% of a radiation length. This amounts to a relative increase in the short-strip module material budget by an estimated 0.8\% due to the kapton layer and 3.9\% due to the SE 4445 layer. At a polar angle of $90$\textdegree, the material budget of the entire ITk material  budget is predicted to be around 50\% of a radiation length \cite{StripTDR}. The new module layers are expected to increase the material budget of entire ITk detector by less than 0.3\% of a radiation length at $90$\textdegree and by less than 1\% overall.

\subsection{Validating the Design}

Two material testing programs were undertaken to inform and validate the interposed module design. The first program established the adhesion of the soft glue to each material involved in the module stack-up, allowing preliminary prototyping to proceed. The second program studied potential evolution of the glue joint strength at the limits of detector conditions, including thermal changes and irradiation to the highest expected doses of the ITk Strip detector.

\subsubsection{Preliminary Material Testing}
\label{sec:proof-of-concept}

During preliminary material tests, the strength of each new glue joint in the module stack-up was measured to verify that the soft glue would adhere sufficiently to all materials involved. The materials of interest included the kapton interposer, the power PCB backing material (a polyamide using kapton HN), and the readout PCB backing material. Shear and peel strengths were measured using samples representing each new glue joint with various test setups and sample geometries.

\paragraph{Preliminary Shear Testing}

\begin{figure}[htbp]
\centering
\includegraphics[width=0.4\textwidth]{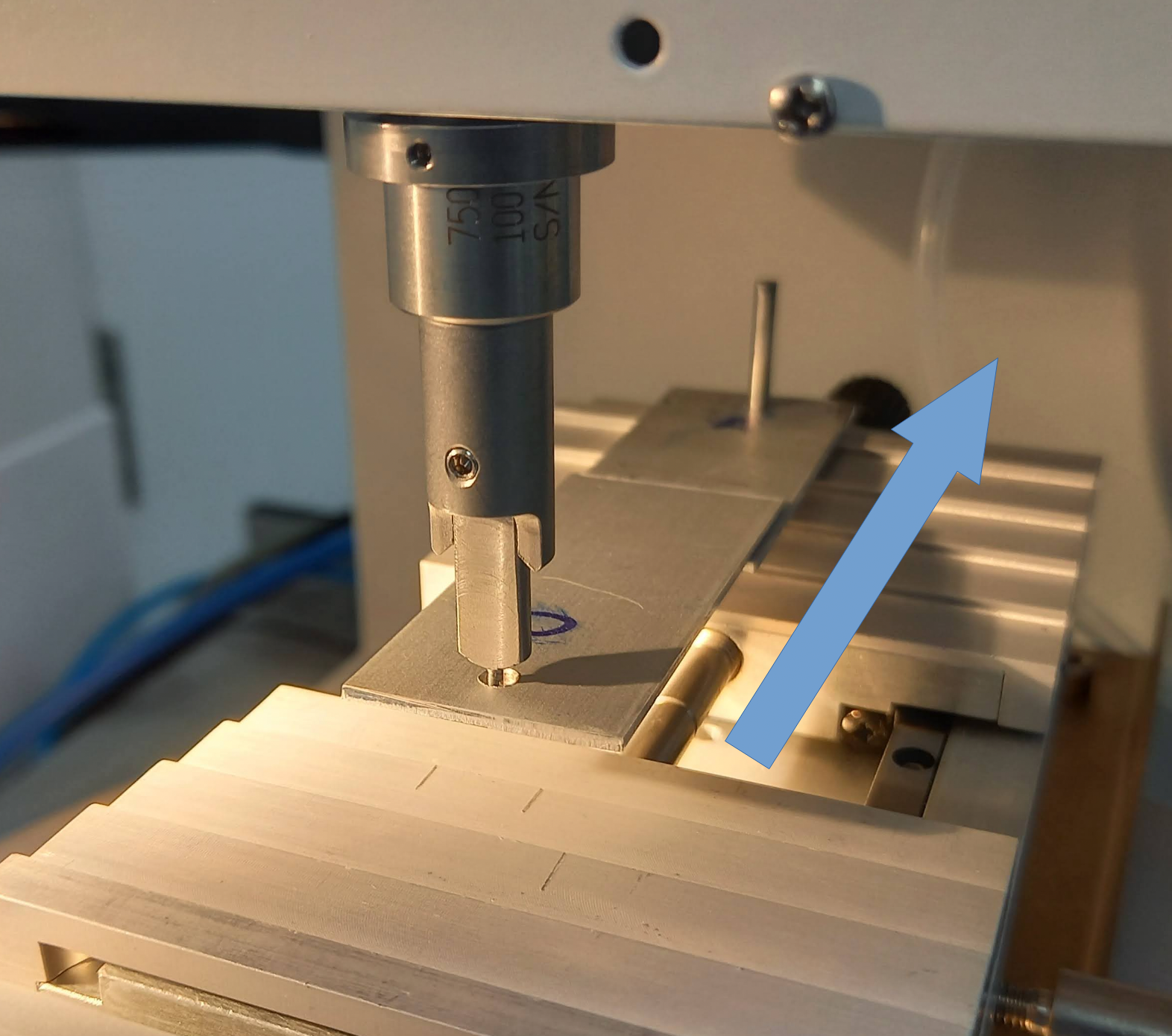}
\caption{The DAGE-4000 bond test system used for shear tests. The sample holder moves along the direction of the arrow while the stationary shear head samples the force.}
\label{fig:dageshear}
\end{figure}

One set of samples was constructed to directly investigate the SE 4445 material properties using aluminum as a reference material. These samples were shear-tested on a DAGE-4000 bond test system, pictured in Figure~\ref{fig:dageshear}. The sample is moved at a fixed speed while a shear test cartridge records the force exerted on the sample with high resolution at a 10 kHz rate. 21 samples were assembled with a stack-up of aluminum--SE4445--aluminum using a 20 x 20 mm lap area to match the working area and travel of the machine.

\begin{figure}[htbp]
\centering
\includegraphics[width=0.9\textwidth]{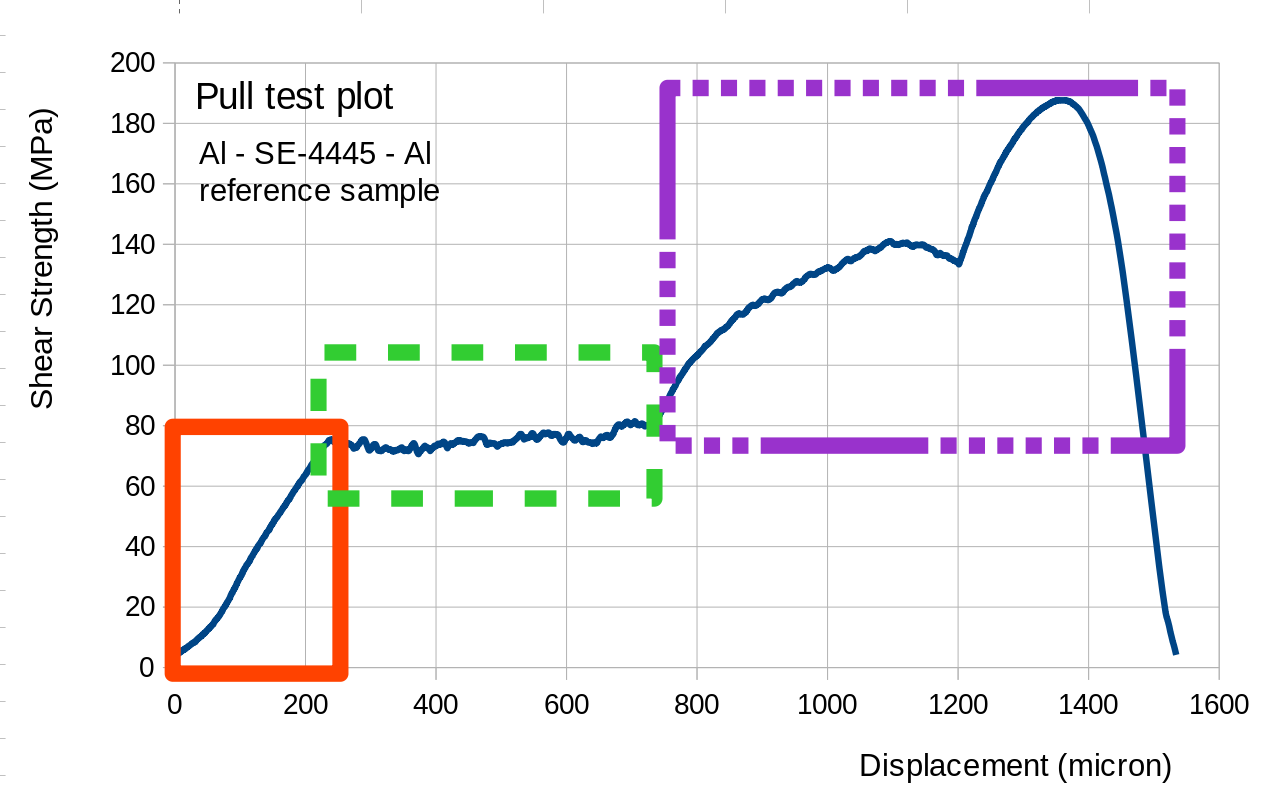}
\caption{Shear test force profile for an aluminum-aluminum lap joint bonded with SE 4445. Three distinct regions are indicated in boxes: the linear elastic region within the red solid outline; the region of lengthening at constant force in the green small-dashed box; the final rupture in the purple box with a varied dashed outline.}
\label{fig:dage_alal}
\end{figure}

A typical force-distance dataset obtained from shear-testing one aluminum--SE4445--aluminum reference sample is plotted in Figure~\ref{fig:dage_alal}. Averaged over 21 samples, a maximum shear strength of 143 $\pm$ 51 kPa was obtained\footnote{Results are reported as average $\pm$ standard deviation. This convention is applied uniformly throughout the paper.}. These results are shown in the first row (marked "DAGE") of Table~\ref{tab:shearandmodulus}. All samples displayed similar behavior: each sample deformed elastically for the first 200 $\mu$m with a modulus of around 30 kPa, followed by a phase of about 400 $\mu$m elongation at 70 kPa shear. Up to that point, the samples displayed very uniform behavior. A final phase is characterized by a large peak marking the sample rupture, where the height of the peak is the main source of the spread in the results obtained. These results provided a basic characterization of the SE 4445 material.

\begin{figure}[htbp]
\centering
\includegraphics[width=\textwidth]{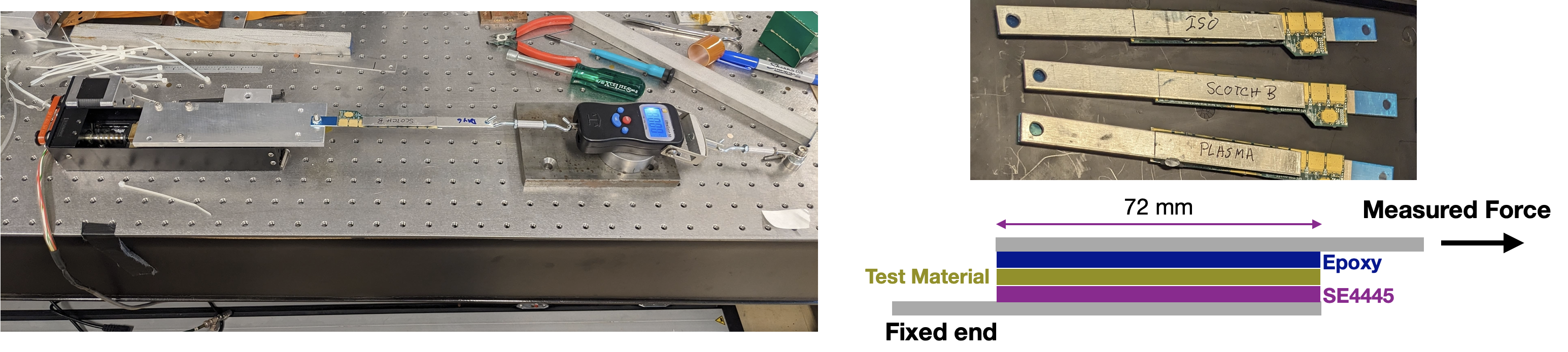}
\caption{The experimental setup used to measure shear strength (left) and the structure of shear test samples (right). Aluminum plates enclose the joint between SE 4445 and the material under test.}
\label{fig:shearsetup}
\end{figure}

A second set of shear tests was carried out at another site using the experimental setup shown in Figure~\ref{fig:shearsetup} to test the adhesion of SE 4445 to other materials. A sample is attached on one end to a digital scale, and on the other end to a plate fastened to a linear stage actuator. The computer-driven actuator pulls the sample away from the digital scale in 1 mm steps. The highest measurement read out by the scale before the sample fails is taken to be the maximum shear strength.

Samples of the structure shown in Figure~\ref{fig:shearsetup} were constructed to represent the various glue joints in the interposed module design. The materials of interest tested in these samples were (1) readout flex PCBs, used to test the adhesion of SE 4445 to the readout PCB backing material, and (2) kapton, used to simultaneously represent the interposer material and the power PCB backing material. The laminated area was chosen to match the size of the power PCB (11 x 72 mm) to demonstrate adhesion to the smallest surface in the interposed module design, as this would be the worst-case scenario for adhesion. 

As the materials of interest were too delicate to attach to the scale and pull without damage, aluminum was used as a rigid reference material on the exterior of the structures. Each sample was constructed by depositing a layer of SE 4445 on aluminum. The aluminum was lightly abraded to ensure that the soft glue would have a stronger bond to the aluminum than to the material of interest, which was held in place on the other aluminum plate using epoxy.

Various surface treatments were applied to the materials of interest before assembling the samples. Readout PCB samples were prepared with treatments of an isopropanol wipe, light abrasion with scotch-brite, or plasma treatment. Kapton samples were prepared with treatments of isopropanol, abrasion with scotch-brite, plasma treatment, or a combination of plasma treatment and scotch-brite. 

As the goal in performing these initial explorations was to test whether the glue joints fail under minimal force, only 3 samples of each material and surface preparation were produced. The samples were pulled in three batches to spot any outstanding evolution of shear strength with the SE 4445 cure time. One sample of each material and surface preparation was pulled after 3, 6, and 10 days of cure time. 

The maximum shear force of all of the samples ranged between $\sim$140-240 N, normalized over the laminated area to a shear strength range of $\sim$177-303 kPa, with no obvious preference for any particular surface preparation or cure time. These shear strengths were deemed sufficient to proceed with prototyping the interposed module design. As surface preparation did not significantly affect the glue joint strength, only a wipe with isopropanol on each surface was incorporated into the prescription for assembling the interposer layer.

\paragraph{Preliminary Peel Testing}

\begin{figure}[htbp]
\centering
\includegraphics[width=0.6\textwidth]{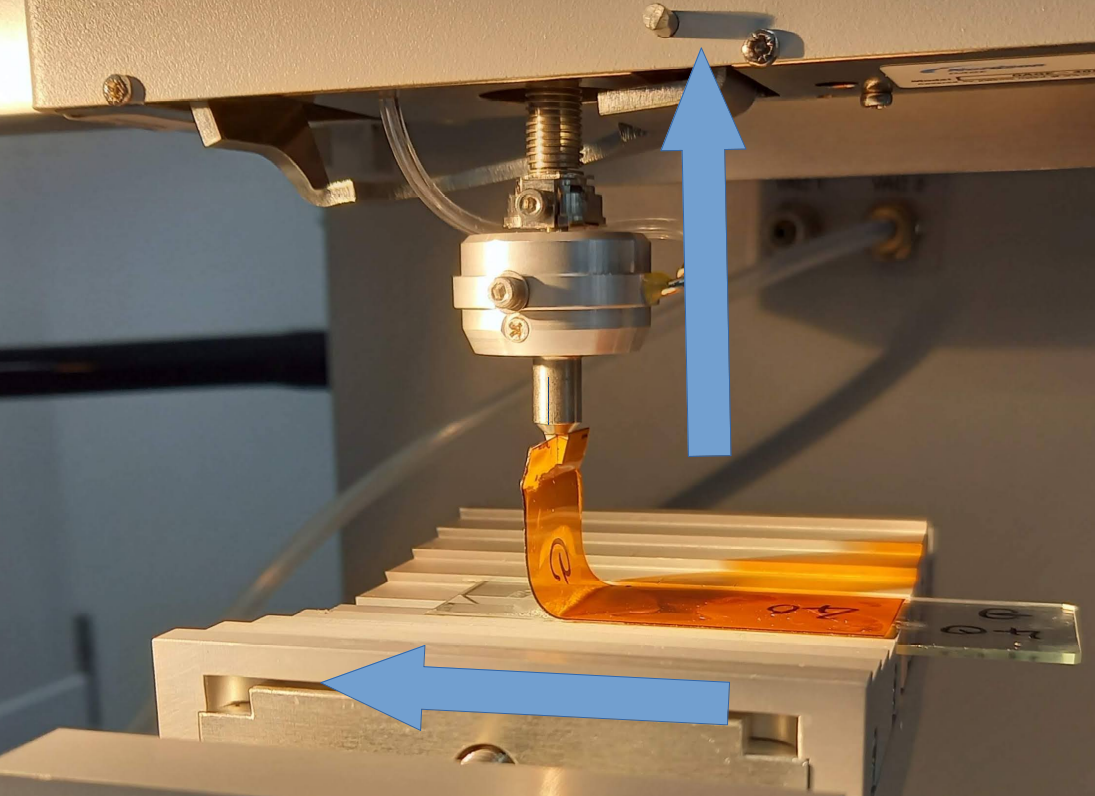}
\caption{The DAGE-4000 bond test system set up for peel tests. The tweezer head moves up while measuring the force, the sample table travels along the arrow at the same speed to conserve the 90\textdegree ~peel angle.}
\label{fig:dagepeel}
\end{figure}

A DAGE-4000 bond test system fitted with a 1 kg tweezer pull cartridge was used to perform a 90\textdegree ~peel test. Figure~\ref{fig:dagepeel} shows a picture of a sample under test. The arrows indicate the direction of travel, with the pull head moving up at the same speed as the sample carrier translates, to ensure force alignment and conserve the 90\textdegree ~pull angle during the test.

Peel tests were carried out on samples of flexible Kapton HN foil glued with SE 4445 to materials of interest to determine peel strengths and assess the uniformity along the glue joint. A 50 mm lap joint length was used to match the maximum travel of the machine. The data was recorded with a 10 kHz pull cartridge force sampling rate.

\begin{figure}[htbp]
\centering
\includegraphics[width=0.9\textwidth]{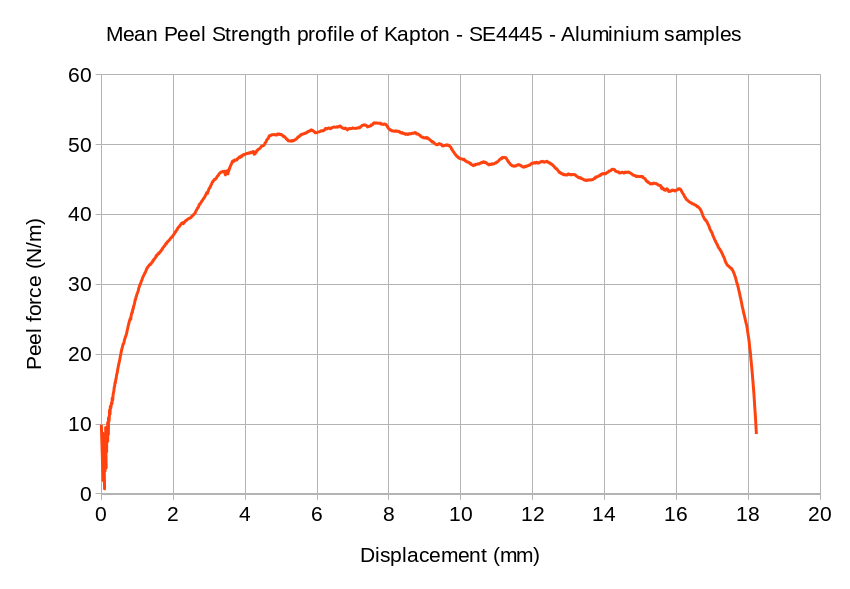}
\caption{Average peel strength profile of 9 samples of kapton--SE4445--aluminum laminate.}
\label{fig:peel_al_ref}
\end{figure}

For a basic assessment of the peel strength of the glue joints in the interposer stack-up, the DAGE-4000 setup was used to test samples of kapton laminated to lightly-abraded aluminum and readout PCB material using SE 4445. The averaged peel strength versus displacement for 9 samples is plotted in Figure~\ref{fig:peel_al_ref}. 

For analysis of the peel test results, the mean and standard deviation of the results were calculated using inter-quartile data to discard the run-in and run-out effects of the peel test. The results are shown in the first two rows (marked "DAGE") of Table~\ref{tab:peel}. This peel strength is markedly weaker than the peel strength of the nominal glue joints incorporating the Loctite Eccobond F 112 epoxy. For comparison, 10 samples of kapton laminated to a glass slide using the epoxy glue yielded a peel strength in excess of 400 Nm$^{-1}$. However, during this initial investigation, the measured peel strength of the SE 4445 glue joints was expected to be sufficient to withstand thermal changes on modules. This assumption is further explored in Section~\ref{sec:QAmaterials}.

\subsubsection{Material Testing at End-of-Life Conditions}
\label{sec:QAmaterials}

A second material testing program was carried out to test for any weakening of the glue joint at the limits of ITk Strip detector conditions. This included shear and peel tests before and after irradiation and/or thermal cycling.

\paragraph{Shear Testing}

\begin{figure}[htbp]
     \centering
     \begin{subfigure}[b]{0.4\textwidth}
         \centering
         \includegraphics[width=\textwidth]{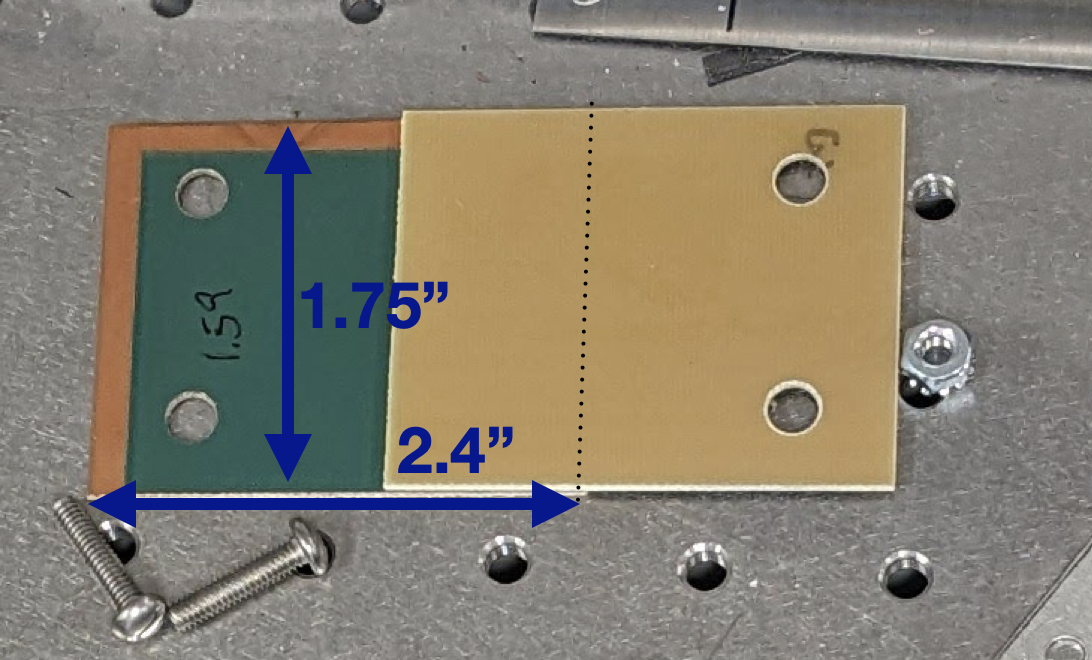}
     \end{subfigure}
     \begin{subfigure}[b]{0.55\textwidth}
         \centering
         \includegraphics[width=\textwidth]{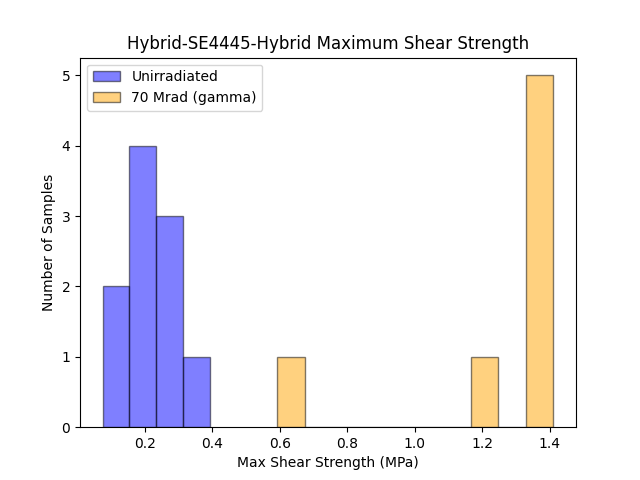}
     \end{subfigure}
\caption{An example readout PCB--SE4445--readout PCB sample demonstrating the dimensions of the samples used for end-of-life testing (left) and the resulting maximum shear strengths of pulled unirradiated and irradiated readout PCB--SE4445--readout PCB samples (right). Four irradiated samples registered a maximum shear strength of $\sim$1.4 MPa, reaching the operation limit of the linear stage actuator, but did not fail.\label{fig:QAshear}}
\end{figure}

40 additional shear testing samples were prepared for this phase of the material testing program: 20 kapton--SE4445--kapton samples and 20 readout PCB--SE4445--readout PCB samples. Readout PCB backing material provided by the readout PCB manufacturer was used to construct these samples. The geometry of these samples, pictured in Figure~\ref{fig:QAshear}, differed from the preliminary testing samples as they were constructed at a different site.

10 samples of each material type were used as control samples and were not irradiated. 10 samples of each type were irradiated at the Sandia Gamma Irradiation Facility using gamma radiation from a Cobalt-60 source to a dose of 70 Mrad, the highest dose expected for any ITk Strip module over its lifetime, including a 1.5x safety factor.

The linear stage actuator experimental setup pictured in Figure~\ref{fig:shearsetup} was used to pull the samples. The resulting maximum shear strength distributions for the readout PCB--SE4445--readout PCB samples are shown in Figure~\ref{fig:QAshear}. The shear force data (N) has been normalized to the laminated area of these samples and is comparable to the shear strengths (kPa) reported in section~\ref{sec:proof-of-concept}. The average shear strength results of the readout PCB--SE4445--readout PCB samples are summarized in Table~\ref{tab:shearandmodulus}.

A clear increase in the shear strength is evident for the irradiated sample population. Only 3 of the 10 irradiated samples could be pulled to failure using this shear testing setup. An additional 4 samples registered maximum shear strengths $\sim$1.4 MPa before the motor of the linear stage actuator reached its limit. The remaining 3 samples were not tested to preserve the motor. None of the irradiated kapton--SE4445--kapton samples could be pulled to failure with the existing setup, likely due to a much larger laminated area compared to the readout PCB--SE4445--readout PCB samples. Weakening of the shear strength of the soft glue joints with irradiation was determined not to be a concern.

However, the increased shear strength of the irradiated samples demonstrates a change in the SE 4445 material that could be reflected in the elastic modulus. This could have troublesome implications for the stress relief provided by the interposer design, which relies on a low-modulus glue layer to decouple thermal stresses. To address this, the elastic modulus of SE 4445 was measured using a tensile test machine to apply small loads to unirradiated SE 4445 samples and SE 4445 samples irradiated to a dose of 70 Mrad \cite{Claybaugh2025}. At a load rate of 25 N/min, the average elastic modulus of unirradiated SE 4445 was found to be 2.7 MPa while the average modulus of the irradiated samples was 15 MPa. By comparison, the typical elastic modulus of an epoxy as used in the original module design is 3-5 GPa. 

\begin{table}[htbp]
\centering
\caption{Shear test and modulus test results demonstrating the change in the SE 4445 properties with irradiation to a dose of 70 Mrad. The modulus of a typical epoxy is shown for comparison. \label{tab:shearandmodulus}}
\smallskip
\begin{tabular}{l r |c c}
\hline
Sample & Treatment& \multicolumn{2}{c}{Shear Strength (N/m)}\\
& & Avg. $\pm$ Std Dev. \\
\hline
Aluminum--SE4445--Aluminum (DAGE) & None & 0.143 $\pm$ 0.051 \\
\hline
Readout PCB--SE4445--Readout PCB & None & 0.197 $\pm$ 0.088 \\
 & Irrad. & > 1.4 \\
\hline
\hline
Material& Treatment&Elastic Modulus (MPa)\\
\hline
SE 4445 & None &2.7 \\
 & Irrad. & 15 \\
 \hline
Typical epoxy & None &3000-5000 \\
 \hline
\end{tabular}
\end{table}

A summary of the modulus results is included in Table~\ref{tab:shearandmodulus}. The SE 4445 modulus increase of less than an order of magnitude is not expected to significantly impact the peak stress on the sensor at end-of-life. A study of the thermal deformation of irradiated modules presented in Section~\ref{sec:stresssection} supports this conclusion. The sensor cracking rate of loaded interposed modules at end-of-life doses will be evaluated empirically with future thermal cycling tests of irradiated fully-loaded staves.

\paragraph{Peel Testing}

80 additional peel testing samples were constructed for further material tests: 40 kapton--SE4445--kapton samples and 40 kapton--SE4445--readout PCB samples. 20 samples of each material type were gamma-irradiated to a total dose of 70 Mrad.

10 irradiated samples of each type and 10 unirradiated samples of each type were then thermal cycled to test whether deformation associated with thermal changes affects peel strength. The kapton--SE4445--kapton samples were attached to aluminum plates to introduce thermal deformation during thermal cycling. The samples underwent 10 cycles between temperatures of $-60$\textdegree C and $+40$\textdegree C, covering the possible range of temperatures expected in the ITk Strip detector ([$20$\textdegree C,$-55$\textdegree C]) and extending the temperature range as allowed by the test setup. 10 samples of each material type were used as control samples and were neither irradiated nor thermal cycled.

A 90\textdegree ~peel test was performed in which the kapton layer was peeled from each sample. The peel strengths are shown in Table~\ref{tab:peel}. As the sample sizes are quite small, with only 10 samples of each type and treatment, and the distribution of peel strengths is quite wide in each case, it is difficult to establish trends of strengthening or weakening glue joints across different treatments. However, within each material type, the average peel strengths across varying treatments are consistent within the standard deviations of the distributions. As a result, it is concluded that thermal cycling and irradiation do not significantly impact the peel strength of the glue joints.

\begin{table}[htbp]
\centering
\caption{Peel strength of kapton--SE4445 glue joint samples for various sample substrates and treatments. Results are reported as average $\pm$ standard deviation. \label{tab:peel}}
\smallskip
\begin{tabular}{l r |c c c}
\hline
Substrate & Treatment & N samples & \multicolumn{2}{c}{Peel Force (N/m)}\\
 & & & Avg. $\pm$ Std Dev. \\
\hline
Aluminum (DAGE) & None & 12 & 49 $\pm$ 8 \\
Readout PCB (DAGE) & None & 7 & 76 $\pm$ 5 \\
\hline
Readout PCB & None & 10 & 91 $\pm$ 17 \\
 & Irrad. only & 10 & 122 $\pm$ 24 \\
 & Thermal cycling only & 10 & 101 $\pm$ 13 \\
 & Irrad. + Thermal cycling & 10 & 93 $\pm$ 21 \\
 \hline
Kapton & None & 10 & 158 $\pm$ 54 \\
 & Irrad. only & 10 & 145 $\pm$ 61 \\
 & Thermal cycling only & 10 & 144 $\pm$ 26 \\
 & Irrad. + Thermal cycling & 10 & 101 $\pm$ 41 \\
\hline
\end{tabular}
\end{table}

\section{Interposer Process Development}
\label{sec:arrayinterposing}

The interposer design has been implemented with the goal of minimizing interruption to existing procedures and infrastructure. Originally, power PCBs were produced and assembled in industry, and readout PCBs were produced in industry and assembled at selected PCB assembly sites. Then, the assembled PCBs were distributed to module assembly sites. To limit disruption to this workflow, the power flex PCBs are laminated with the soft glue and kapton layers in industry, and the readout flex PCBs are laminated at the site that receives readout PCB flexes from industry and distributes them to readout PCB assembly sites. Then, the laminated flex PCBs are fed back into the original production workflow. Module assembly proceeds using the established tooling and procedures, with the addition of shims underneath the PCB placement tooling to account for the extra thickness of the laminated flex PCBs.

Over 25,000 flex PCBs must be laminated to meet the needs of module production \cite{StripTDR}. Readout and power flex PCBs are manufactured in arrays, with 10 power flex PCBs per power flex PCB array and 7 readout flex PCBs per array. To efficiently modify large numbers of flex PCBs at production scales, the entire array of flex PCBs is laminated prior to cutting out the individual flex PCBs from the arrays. The lamination step occurs after loading Surface Mount Devices (SMDs), as the temperature during reflow soldering exceeds the maximum operating temperature of SE 4445 \cite{SE4445}, and prior to ASIC-loading to simplify the lamination processes and prevent damage to ASICs.
 
SMD-loaded readout flex PCB arrays are relatively flat and weight can be applied to most of the array surface area while curing the soft glue. However, an SMD-loaded power flex PCB is crowded with components such as capacitors and coils that extend several millimeters above the surface of the flex. Care must be taken to apply weight uniformly to the array while avoiding damage to sensitive components. Separate lamination procedures have been developed for readout flex PCB arrays and for power flex PCB arrays.

\subsection{Process to Laminate Power Flex PCBs} 

The process to laminate an array of power flex PCBs was developed at Lawrence Berkeley National Laboratory, the site responsible for testing and distributing power PCBs, before being transferred to an industry partner where it will be carried out during production. 

\begin{figure}[htbp]
\centering
\includegraphics[width=\textwidth]{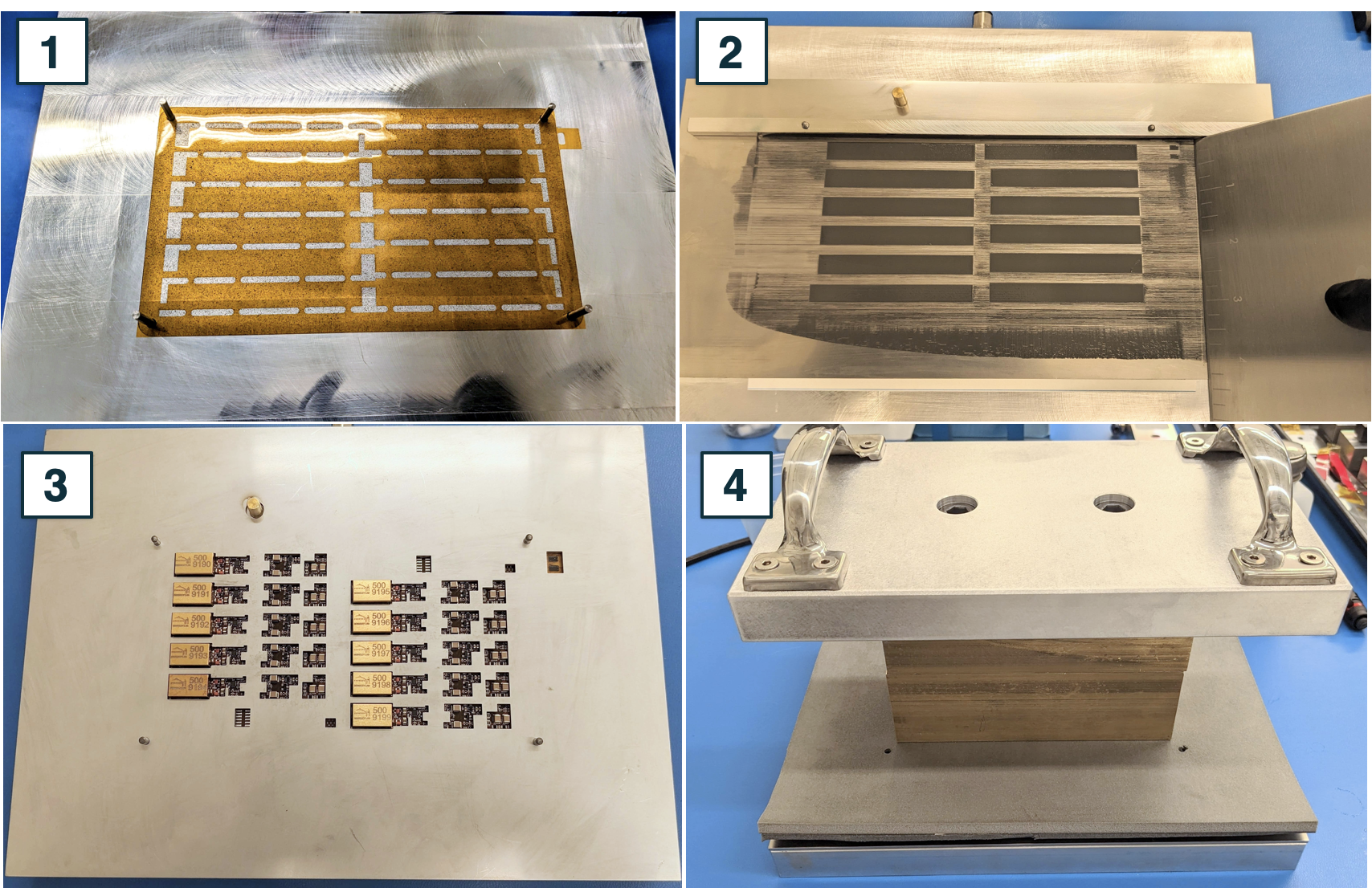}
\caption{Key steps of the process developed to laminate power flex PCB arrays. An explanation of each step is given in the text \cite{AnneInterposingProc}.}
\label{fig:PBinterposingsteps}
\end{figure}

Figure~\ref{fig:PBinterposingsteps} demonstrates the power PCB array-laminating process. In Step 1 of Figure~\ref{fig:PBinterposingsteps}, a piece of kapton laser-cut into the shape of the power flex PCB array is placed on a vacuum tool and aligned using dowel pins. A metal foam insert in the vacuum plate diffuses the vacuum, holding the kapton in place without dimpling it. Next, in Step 2, a 150-$\mu$m-thick stencil is placed over the kapton. A stainless steel glue applicator is used to deposit the SE 4445 glue over the stencil windows. The stencil is lifted from the kapton, and the flex PCB array is placed over the kapton using dowel pins in the corners for alignment. 

\begin{figure}[htbp]
\centering
\includegraphics[width=\textwidth]{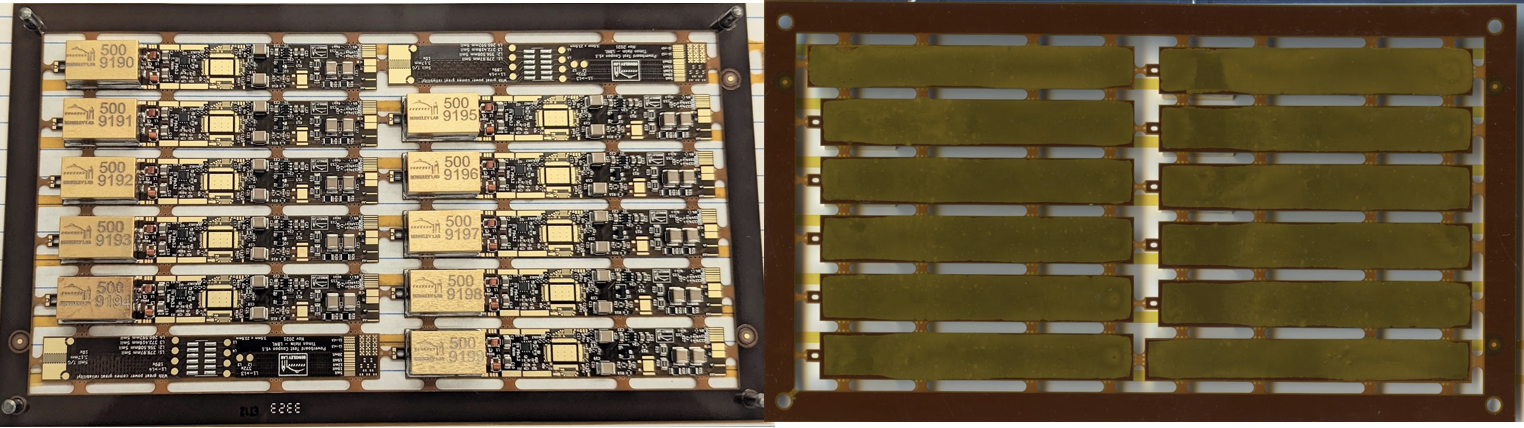}
\caption{Front and back views of a successfully laminated power flex PCB array. A layer of SE 4445, protected by a layer of kapton, has been added to the back of the original array \cite{AnneInterposingProc}.}
\label{fig:PBarray}
\end{figure}

The minimum thickness of the glue layer is set using glass beads of diameter 106 $\mu$m $\pm$ 6\% \cite{Glassbeads} mixed into the glue. Weight must be applied to as much of the flex PCB array surface as possible to ensure a uniform glue layer thickness. To this end, a stainless steel mask was designed with cutouts such that pressure can be applied to the flex PCBs at many points without damaging sensitive components. Step 3 in Figure~\ref{fig:PBinterposingsteps} shows the mask laid over the array. In Step 4, a layer of foam covers the mask and array to protect the components, and a 5-kg block of brass is set on top of the foam. The glue is allowed to cure for at least 4 hours before the weight is removed. A successfully laminated power flex PCB array is shown in Figure~\ref{fig:PBarray}.

In evaluating the quality of a laminated array, the primary considerations are (1) achieving a uniform glue thickness across the array, (2) providing sufficient glue coverage underneath the flex PCB wirebonding pads to support wirebonding, and (3) preventing residue from the SE 4445 from contaminating the wirebonding pads.

Over the first 10 arrays laminated with the finalized procedure and tooling, the average glue thickness was found to be 109 $\pm$ 10 $\mu$m, meeting the target glue layer thickness. Sufficient glue coverage on the underside of the flex PCBs is achieved through the design of the stencil windows, which deposit glue to cover the back of each flex PCB within 1 mm of the edge. The glue coverage then increases during compression. This yields >80\% glue coverage on each flex PCB, which has been empirically demonstrated to be sufficient to support wirebonding on the top of the flex PCB while minimizing the risk of glue contamination spreading to wirebonding pads.

Glue contamination on wirebonding pads is spotted through visual inspection by the lamination process operator after curing. Oily residue can seep from the SE 4445 material and contaminate the wirebonding pads, preventing effective wirebonding. If the contamination is spotted within several days of lamination, the residue can be removed from wirebonding pads using acetone, and full wirebondability is recovered. Wirebondability is checked with a wirebond pull test, wherein test bonds are attached to wirebonding pads on several representative flex PCBs and pulled until they fail. Test bonds are required to meet specifications on minimum wirebond load, average wirebond strength, and mode of failure. Of the first 10 arrays laminated with the finalized procedure, all arrays passed the wirebond pull test after routine cleaning.

Power flex PCBs are then used to complete power PCB assembly with ASIC attachment and wirebonding. During production, 150 power PCBs will be laminated and assembled in industry per week. After passing previously-established electrical quality control testing \cite{Haber_2023} at Lawrence Berkeley National Laboratory, the laminated power PCBs are distributed to module assembly sites.

\subsection{Process to Laminate Readout Flex PCBs} 

The process to laminate readout flex PCBs was developed at Rutherford Appleton Laboratory and transferred to Liverpool University, where it will be performed at production scales. Liverpool University was chosen as the lamination site to limit shipping costs, as readout flex PCBs are delivered and tested at this site before distribution to module assembly sites. The target rate for production is lamination of seven readout flex PCB arrays per day. 

\begin{figure}[htbp]
  \centering
  \includegraphics[width=\textwidth]{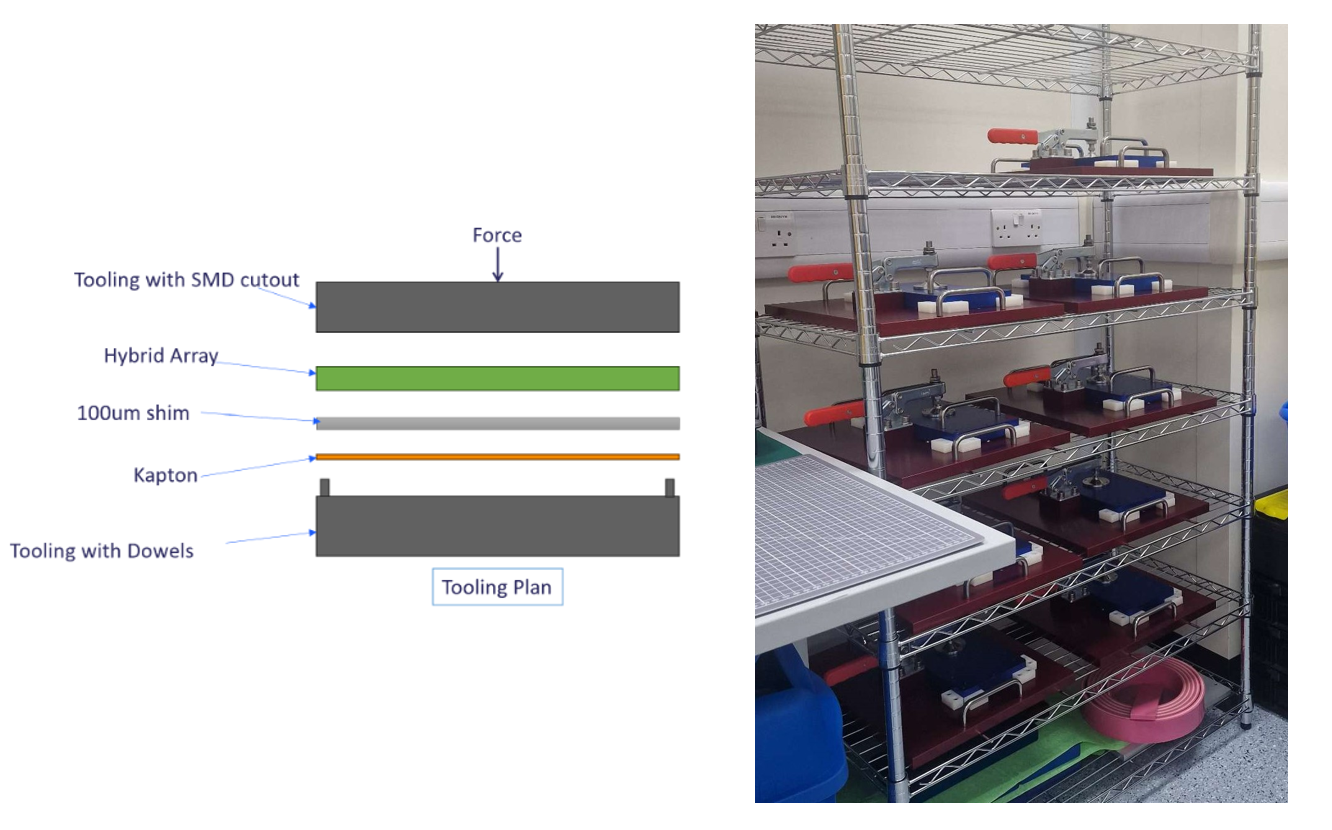}
  \caption{Left: Schematic demonstrating readout flex PCB lamination. Right: Production tooling at Liverpool University allowing up to 10 lamination operations per day.}
  \label{fig:toolLiverpool}
  \end{figure}

\begin{figure}[htbp]
  \centering
  \includegraphics[width=\textwidth]{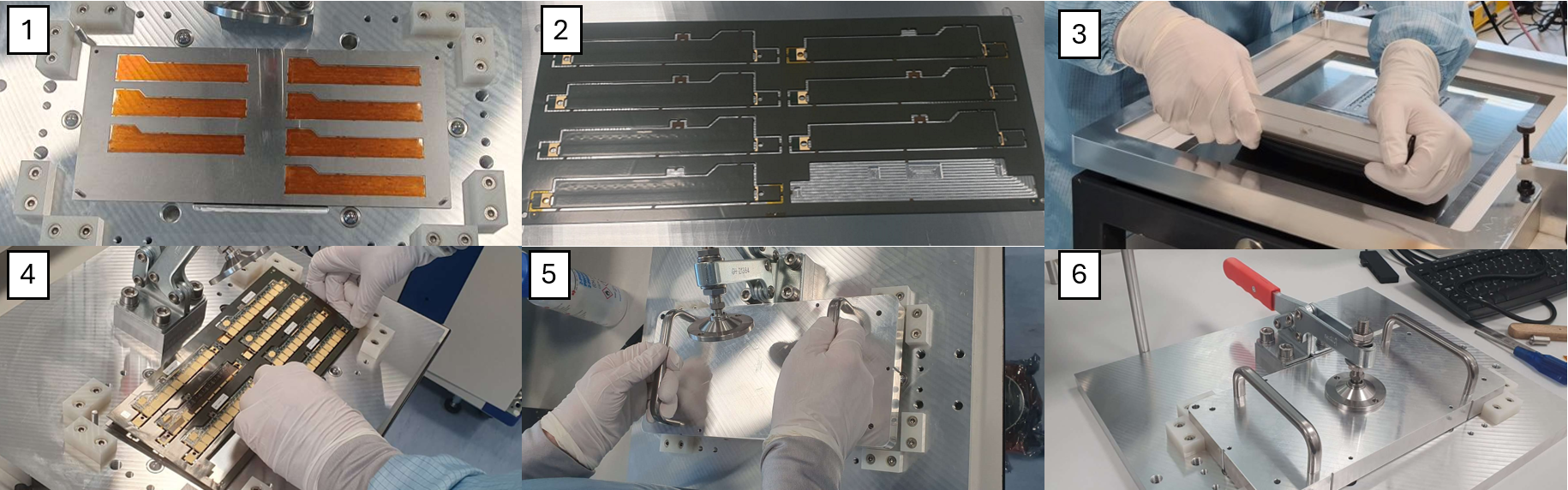}
  \caption{A photographic guide to the steps to laminate readout flex PCBs. An explanation of each step is given in the text. }
  \label{fig:hyInter}
  \end{figure}

Figure~\ref{fig:toolLiverpool} shows the tooling developed for the readout PCB lamination process. The steps of the lamination process are demonstrated in Figure~\ref{fig:hyInter}. First, kapton pre-cut to the shape of the readout flex PCBs is aligned using dowel pins on the tooling base. A shim of 100 $\mu$m thickness is aligned on top of the kapton. 


In Step 2 of Figure~\ref{fig:hyInter}, the readout flex PCB array is aligned face-down on a second tooling plate that has cutouts to accommodate the SMD components. The array is held in place via dowel pins machined to a length of 500 $\mu$m. In Step 3, a stencil is aligned on the array and a squeegee is used to deposit SE 4445 over the stencil image onto the back of the array. Next, in Step 4, the array is removed from the second tooling plate and aligned face-up on the first tooling plate on top of the kapton and the shim. Finally, in Steps 5 and 6, the second tooling plate is placed over the array and clamped down while the SE 4445 cures. Unlike the power flex PCB lamination process, no glass beads are used in the SE 4445. Instead, the shim sets the SE 4445 layer thickness during the compression step.

\begin{figure}[htbp]
  \centering
  \includegraphics[width=\textwidth]{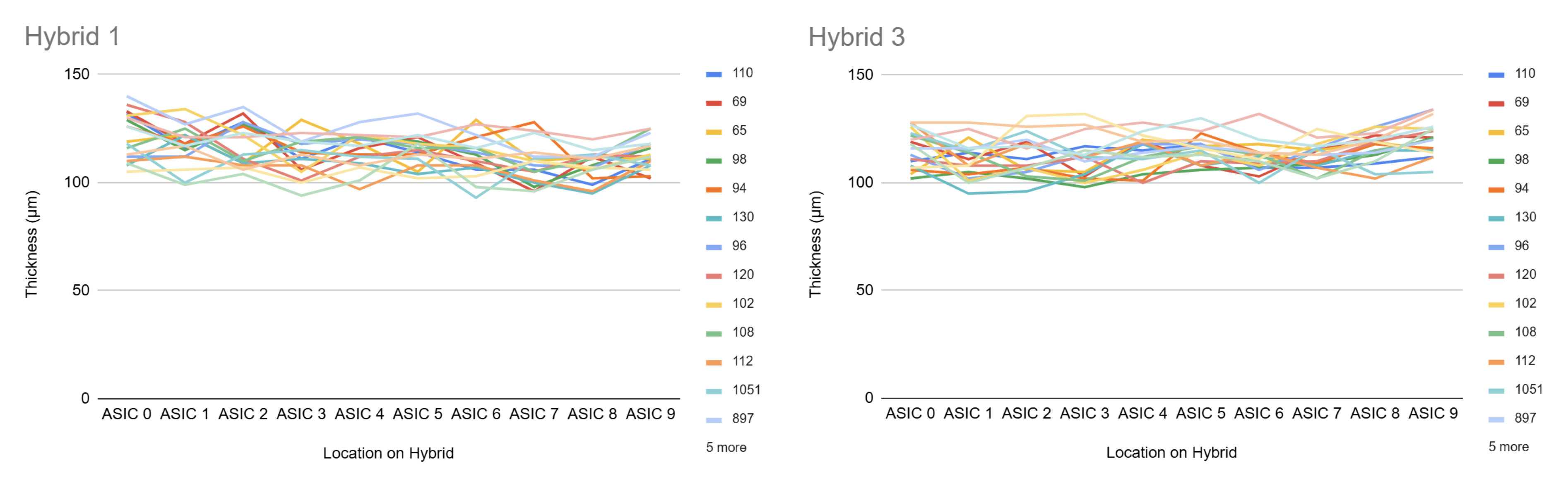}
  \caption{Measurements of SE 4445 thickness from 28 arrays across readout flex PCBs in two different array positions. The average thickness is 115 $\pm$ 9 $\mu$m.}
  \label{fig:hybridthickness}
  \end{figure}

As in the power flex PCB lamination process, the primary considerations for quality control are uniform glue thickness, sufficient glue coverage to the edge of the flex PCBs to support wirebonding, and cleanliness of wirebonding pads. Figure~\ref{fig:hybridthickness} shows the SE 4445 glue thickness measured at 10 points on the readout flex PCBs in the first and third flex PCB positions across a total of 28 arrays. The average thickness across all these points was found to be 115 $\pm$ 9 $\mu$m, demonstrating sufficient uniformity of SE 4445 thickness and consistency across many arrays.

During process development, there were initial concerns of insufficient SE 4445 coverage on the back of the flex PCBs and cleanliness of wirebonding pads. SE 4445 coverage has been improved over time by modifying the stencil openings. Concerns about SE 4445 residue contaminating wirebonding pads have been addressed by implementing cleaning steps and a strict visual inspection check on outgoing laminated arrays.

\section{Prototype Interposed Modules}
\label{sec:prototypesection}

\subsection{Design Validation Results}
\label{sec:QA}

Prototype interposed modules were constructed to investigate whether interposed modules meet ITk Strips requirements under the extreme conditions a module may endure while installed in the ATLAS detector. These tests evaluate the impact of the interposer design on module thermal performance, tracking performance in a test beam environment, long-term reliability, and electrical performance at extra-cold temperatures. In addition, the impact of interposing on radiation hardness is evaluated by studying module performance after irradiation to total expected doses, including a 1.5x safety factor. Overall, no alarming issues were spotted during design validation testing.

As the production lamination methods described in section~\ref{sec:arrayinterposing} were not yet developed during this investigation phase, these prototype modules were constructed using existing readout PCBs and power PCBs that already had ASICs assembled and had been wirebonded. These already-assembled PCBs were individually hand-laminated at various sites. Hand-lamination methods varied from site to site. An example procedure is described in \cite{Liu:2938335}.

\paragraph{Thermal Performance}
The impact of interposing on module cooling was evaluated, as the interposed module design adds new layers to the module stack-up, including 50 $\mu$m of thermally-insulating kapton. A comparison of the thermal performance of unirradiated modules, both interposed and non-interposed, is further discussed in Section~\ref{sec:QC} and is shown to differ by a few degrees Celsius, consistent with the prediction of a thermal simulation \cite{Beck:2020qme}.

A short-strip module that was gamma-irradiated to a total dose of 70 Mrad using a Cobalt-60 source was used to evaluate the impact of irradiation on thermal performance. The negative-temperature coefficient (NTC) thermistor readings from the readout PCBs and power PCB were compared before and after irradiation while powering the module and maintaining its environment at a temperature of $20$\textdegree C. The NTC readings before and after irradiation differed by less than $1$\textdegree C, demonstrating negligible impact of irradiation on the thermal performance of interposed modules.

\paragraph{Test Beam Performance}

The impact of interposing on the performance of a module under beam conditions was evaluated during two test beam campaigns. The first test beam campaign occurred at the DESY II test beam facility \cite{DIENER2019265} in Hamburg, Germany during June of 2024 with a beam composed of 5 GeV electrons. Two interposed barrel modules, including one long-strip module and one short-strip module, were tested.

\begin{figure}[htbp]
     \centering
     \begin{subfigure}{0.85\textwidth}
         \centering
         \includegraphics[width=\textwidth]{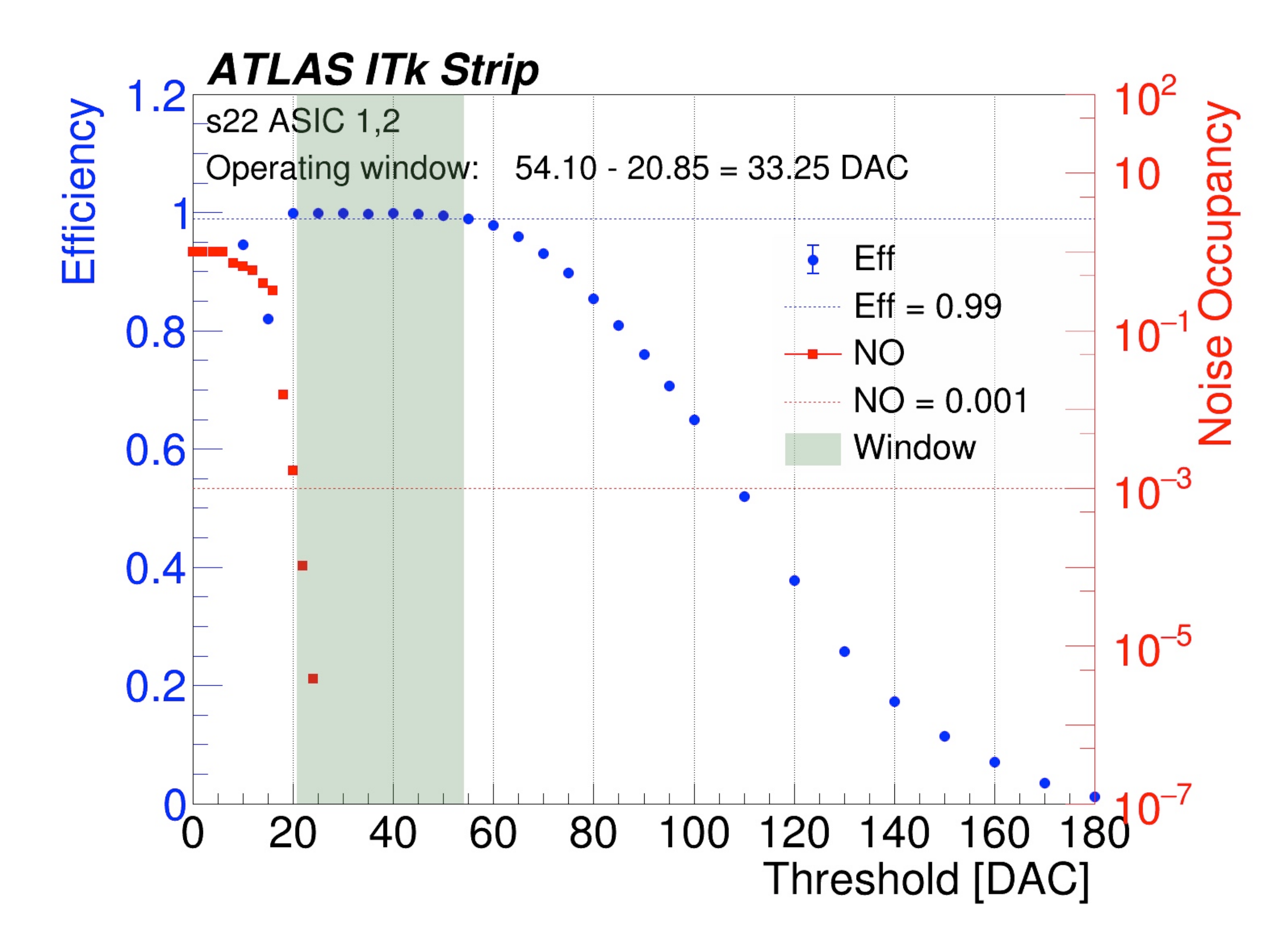}
         \caption{}
         \label{fig:unirradiatedSS}
     \end{subfigure}
     \begin{subfigure}{0.85\textwidth}
         \centering
         \includegraphics[width=\textwidth]{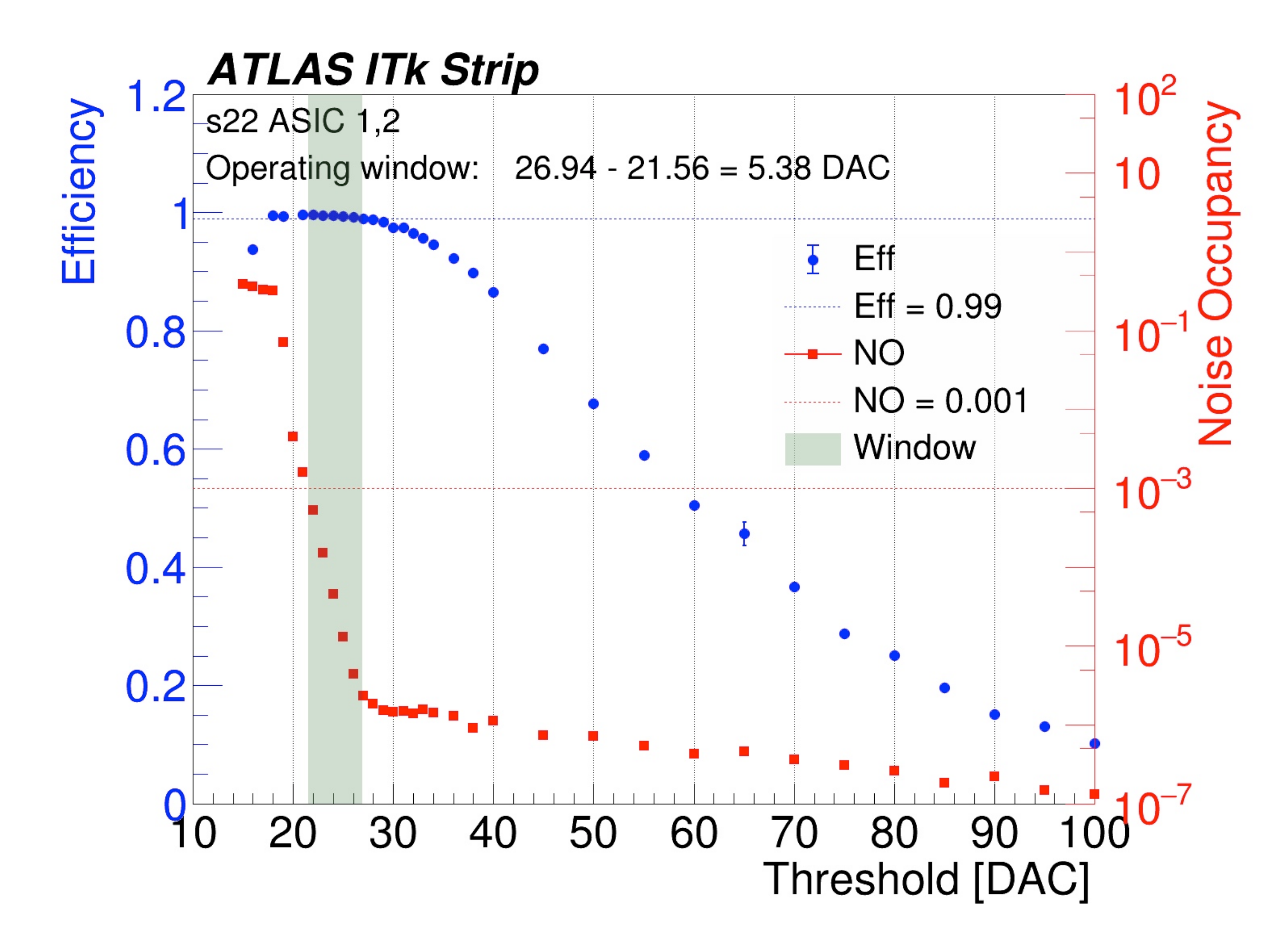}
         \caption{}
         \label{fig:irradiatedSS}
     \end{subfigure}
\caption{An example operating window (shaded green) for one column of strips in a short-strip module (a) before and (b) after proton-irradiation to a total dose of 54.4 Mrad.}
\label{fig:testbeamplots}
\end{figure}

To successfully operate an ITk Strip module in beam conditions according to specifications, a range of thresholds (called an \textit{operating window}) must be found that permits an efficiency of over 99\% with a noise occupancy of less than 0.1\%. Figure~\ref{fig:unirradiatedSS} highlights the operating window in green that satisfies this specification for an example set of readout channels on the short-strip module. Operating windows that meet ITk Strip specifications were found on both the long-strip and short-strip modules.

The short-strip module was then irradiated at the CERN IRRAD facility to a total integrated dose of 54.4 Mrad with a proton fluence of $1.8$x$10^{15}$ protons/cm$^2$. A second test beam campaign held at the CERN SPS test beam facility in September 2024 tested the performance of this irradiated short-strip module using a beam of 120 GeV protons. The operating window is expected to narrow with irradiation due to reduced charge collection efficiency and increased noise caused by radiation damage. Figure~\ref{fig:irradiatedSS} shows the operating window of the irradiated module for a sample group of readout channels. 

The irradiated module was tested in an asynchronous data-taking mode with a known systemic decrease in performance due to non-constant delays between particle arrival and sampling in the module. However, even under these conditions, operating windows were successfully located in 3 out of 4 readout channel groups. Other modules tested under the same asynchronous conditions displayed similar readout issues, including narrow operating windows, but were later retested in synchronous data-taking mode, where operating windows were shown to significantly increase in all readout channel groups. As this known data acquisition issue appears in the asynchronous mode regardless of whether a module is interposed, it is expected that the irradiated interposed module retested with synchronous data-taking will show similar improvement, although analysis of the retested data has not been completed at the time of writing. Overall, no performance reduction in interposed modules compared to non-interposed modules tested under the same conditions was observed.

\paragraph{Long-Term Reliability}

An ITk Strip module will need to withstand many changes of temperature over its expected 10-year lifetime, including yearly planned detector shutdowns and power failure events. To ensure modules will survive these conditions, a thermal cycling program of 10 cycles between $-35$\textdegree C $\rightarrow$ $+20$\textdegree C $\rightarrow$ $-35$\textdegree C is performed on every module as part of the standard quality control procedure. Additionally, a stress test of over 100 thermal cycles had been performed on a handful of early ITk strip modules to understand the long-term reliability of module performance \cite{Abe_moduleTC}. 

A similar extended thermal cycling test was carried out using prototype interposed modules to demonstrate any impact of interposing on long-term reliability. Two interposed modules, one long-strip and one short-strip, underwent an extensive thermal cycling program over a period of several months: 10 cycles between temperatures of $-35$\textdegree C and $+20$\textdegree C, followed by 190 cycles between a lower temperature of $-44$\textdegree C and an upper temperature that ranged from $+20$\textdegree C to $+56$\textdegree C. 

The electrical performance, including noise performance and sensor current as a function of high voltage, was found to be consistent for both modules before and after extended thermal cycling in the temperature range $[+40, -44]^\circ$C. During the last few cycles between $[+56, -44]^\circ$C, the long-strip module developed a sharp rise in sensor current at a bias voltage of $-$320 V and could not be biased to the nominal bias voltage of $-$350 V, an issue known as \textit{early high voltage breakdown} \cite{Helling:2019zea}. While early high voltage breakdown can arise as an electrical manifestation of thermal stress, it did not appear in this case until nearly 200 thermal cycles to more extreme upper temperatures than expected during detector operation. Further details of this study can be found in \cite{Liu:2938335}.

\paragraph{Extra-Cold Testing}

The standard module quality control testing procedures include thermal cycling in a range of $+20$\textdegree C and $-35$\textdegree C, which is the coldest expected operating temperature for ITk Strip modules. As mentioned in Section~\ref{sec:intro}, in cases of power or cooling failures, the detector may experience temperatures as low as $-55$\textdegree C. Thermal cycles involving colder-than-quality-control temperatures were performed on a selection of interposed modules to check whether electrical problems appear below $-35$\textdegree C. 

\begin{figure}[htbp]
     \centering
     \begin{subfigure}{0.99\textwidth}
         \centering
         \includegraphics[width=\textwidth]{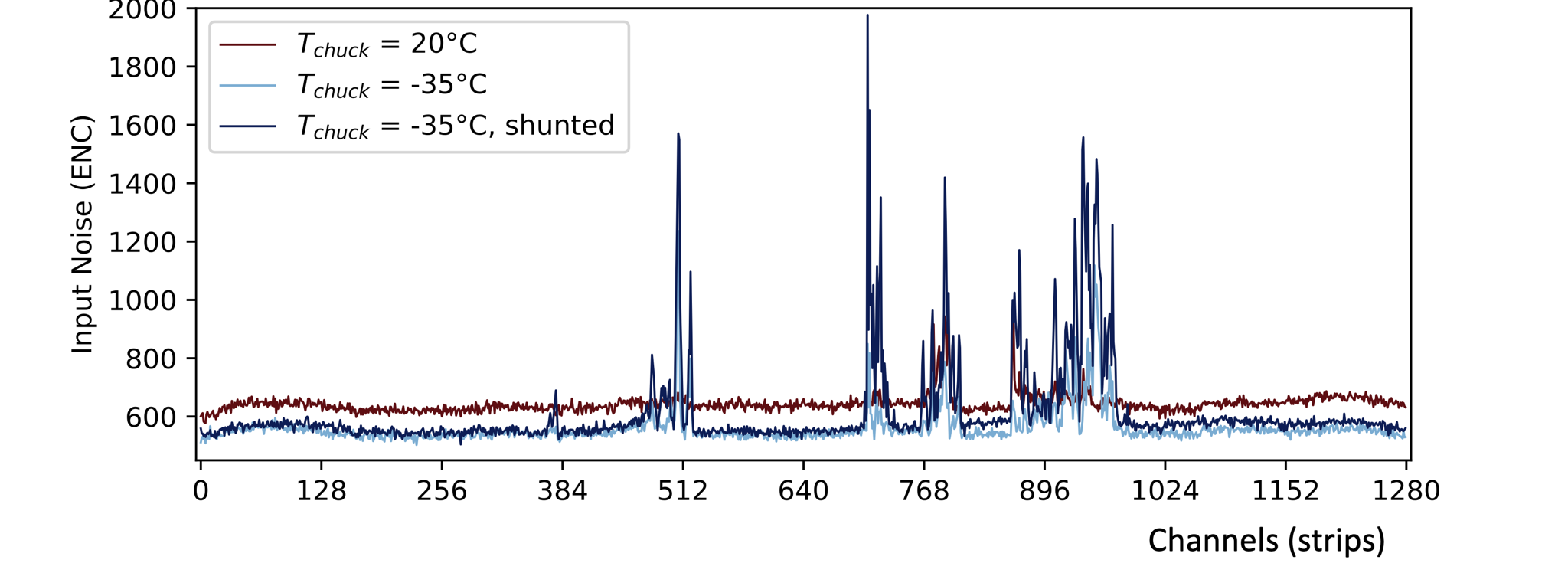}
         \caption{}
         \label{fig:CN_typical}
     \end{subfigure}
     \begin{subfigure}{0.9\textwidth}
         \centering
         \includegraphics[width=\textwidth]{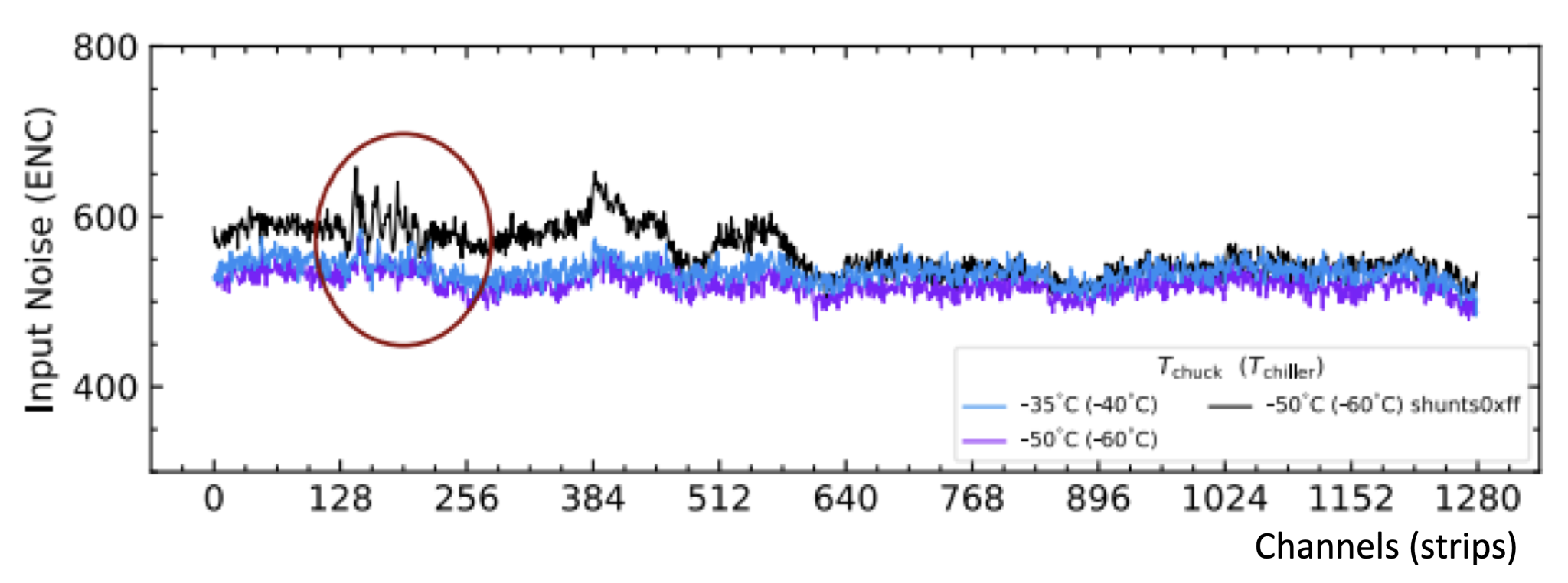}
         \caption{}
         \label{fig:CN_trueblue}
     \end{subfigure}
\caption{Input noise in units of Equivalent Noise Charge (ENC) for each readout channel in a group of strips on a sensor, where each strip corresponds to one readout channel. In the absence of anomalous noise issues, the input noise should be relatively uniform across channels, decreasing slightly when the module is tested at colder temperatures. (a) Typical cold noise behavior (excess noise spikes while testing cold) of a long-strip module appearing at $-35$\textdegree C prior to the implementation of any cold noise mitigation strategies. (b) Mild cold noise remaining on some readout channels of a short-strip module assembled using Loctite Eccobond F 112 epoxy as a cold noise mitigation strategy. Mild cold noise appearing at environmental temperatures of $-35$\textdegree C (blue curve) and $-50$\textdegree C (purple) is exacerbated when enabling shunts on the readout ASICs that increase the current consumption (black).}
\label{fig:CNplots}
\end{figure}

Before the sensor cracking problem was detected, the most critical issue facing ITk Strip barrel modules was that of \textit{cold noise}, detailed in \cite{IanCNProc}. In summary, cold noise is excess noise that tends to appear on readout channels at cold temperatures, often visible at $-35$\textdegree C. Figure~\ref{fig:CN_typical} shows an example of cold noise, which appears as spikes in the input noise on some readout channels. The source of cold noise was determined to be the mechanical vibration of multi-layer ceramic capacitors on the power PCB inducing noise on the strips. After a thorough investigation, the chosen mitigation strategy was the replacement of a previous epoxy between the flexes and the sensor with the current Loctite Eccobond F 112 epoxy \cite{TrueBlue} in the module stack-up. This new epoxy significantly reduced cold noise. However, slight cold noise remained on some short-strip modules, as shown in Figure~\ref{fig:CN_trueblue}. 

As mentioned earlier in Section~\ref{sec:solutionsection}, test modules had been assembled using SE 4445 instead of the epoxy between the flex PCBs and sensor as an attempted cold noise mitigation. No cold noise was observed on these modules. However, electrical issues such as high sensor currents were observed, and SE 4445 was excluded as a replacement glue. These previous cold noise studies indicated that the addition of SE 4445 in the interposed module stack-up could have a mitigating effect on cold noise.

It was therefore unsurprising that cold noise did not appear on any prototype interposed module in a sample of 79 interposed long-strip modules and 9 interposed short-strip modules tested at $-35$\textdegree C. Of these, 7 long-strip modules and 3 short-strip modules were tested at a colder temperature of $-50$\textdegree C and did not exhibit cold noise. Cold noise also did not appear during or after 10 thermal cycles between $-50$\textdegree C $\rightarrow$ $+20$\textdegree C $\rightarrow$ $-50$\textdegree C. 

\begin{figure}[htbp]
\centering
\includegraphics[width=0.8\textwidth]{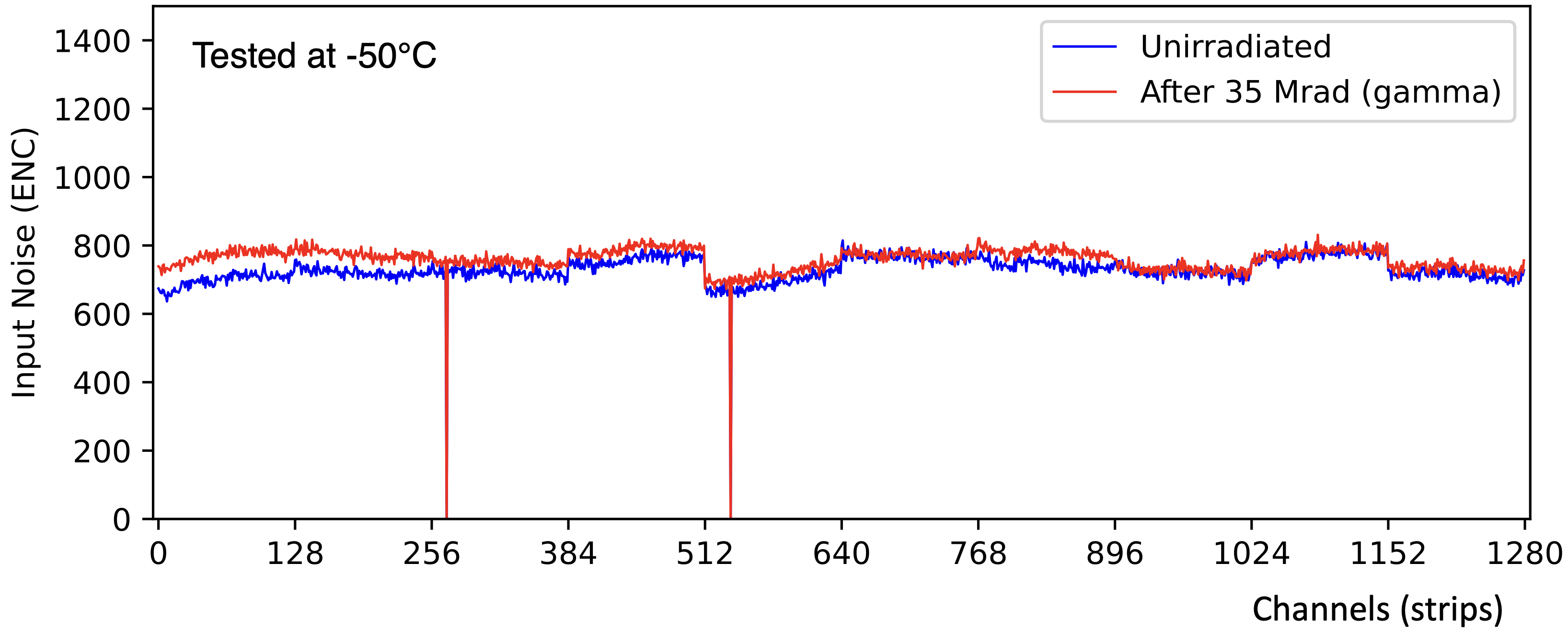}
\caption{Noise as a function of readout channel for 1280 strips, while testing at a temperature of $-50$\textdegree C before and after gamma irradiation to a dose of 35 Mrad. No cold noise is observed. Two readout channels on these readout ASICs were known to be unresponsive prior to module assembly.}
\label{fig:CNirradiated}
\end{figure}

Furthermore, cold noise did not appear after irradiation to total expected ITk Strip doses (including a 1.5x safety factor). One interposed long-strip module and one interposed short-strip module were irradiated to a dose of 35 Mrad and 70 Mrad respectively with gamma radiation from a Cobalt-60 source. No cold noise appeared on either module before or after irradiation. Figure~\ref{fig:CNirradiated} shows a comparison of noise before and after irradiation for an example set of readout channels on the long-strip module. While the overall noise rises slightly after irradiation as expected, no cold noise appears, even while testing at a temperature of $-50$\textdegree C.

Considering the results from this sample of prototype interposed modules, a welcome effect of the interposer design appears to be the elimination of cold noise in the temperature range relevant to operation in ATLAS. This also suggests a link between cold noise and sensor stress.

\subsection{Quality Control Results}
\label{sec:QC}

Given these promising results, a second phase of prototyping began with the goal of producing a sufficient number of long-strip interposed modules to populate several prototype interposed staves. These interposed modules were constructed with readout PCBs and power PCBs that were interposed using either hand-lamination methods or array-lamination methods as array-laminated parts became available. 

These modules were subjected to the complete mechanical and electrical quality control procedures previously defined for ITk Strip modules described in \cite{AbeQCProc,ABC130paper}. The quality control results of 79 long-strip interposed modules are examined in this section. Overall, these results do not reveal major differences in the electrical performance of interposed modules relative to non-interposed modules.

\paragraph{Mechanical Quality Control}

Mechanical quality control tests are routinely carried out on each module to assess the quality of the module assembly. These tests include measurements of the flatness of the sensor of the assembled module, positioning of the flex PCBs on the sensor, and height of the flex PCBs, from which the height of the epoxy layer beneath each PCB is deduced. Modules must also satisfy requirements on the amount of failed and reworked wirebonds.

Module height specifications were modified as a result of the new design. The total package height of interposed modules increased by $\sim$150 $\mu$m due to the new layers. The clearance needed to avoid collisions with ITk infrastructure during installation was re-evaluated, resulting in a new total package height limit of 6110 $\mu$m, which was satisfied by all interposed modules. 

The failure rate of wirebonds could in theory be affected by any SE 4445 residue on wirebond pads or lack of SE 4445 glue coverage supporting wirebonding pads missed by quality control of laminated parts. However, all interposed modules passed the wirebond quality control specifications in a sample size of 56 modules for which data was available. No changes were made to the specifications, and no significant increase in wirebond failures or reworked bonds was found compared to non-interposed modules. 

The interposed modules also passed requirements on the flatness of the sensor in the assembled module. Sensor deformation, which is measured at room temperature, can indicate stress acquired during module assembly and can interfere with loading tooling when attaching modules to staves. Of the 35 modules where data was available, all 35 interposed modules passed the existing specifications on module flatness.

The overall mechanical quality control pass rate of interposed modules was 85\%. Failures tended to occur in the areas where changes were made to the module assembly process. In the original module assembly procedure, the sensor is held flat with a vacuum tool. Each flex PCB is placed onto the sensor using specialized tooling that positions the PCB above the sensor at a height such that the epoxy layer between the flex PCB and the sensor compresses to a nominal thickness of 120 $\mu$m.

However, the laminated flex PCBs are thicker than the bare flex PCBs by roughly 150 $\mu$m due to the thickness of the kapton and SE 4445 layers. Therefore, to maintain the correct epoxy layer thickness, the flex PCB must be placed around 150 $\mu$m farther from the sensor than the original tooling allows. To account for this, shims are placed on the module assembly vacuum tooling to elevate the flex PCB placement tooling by the correct amount.

Several early interposed modules built with incorrect shim heights were failed due to the epoxy height falling outside of the epoxy layer thickness specification. If the shims are too thin, the laminated PCB is assembled too close to the sensor. The epoxy can squeeze out beyond the flex PCB, which can prevent wirebonding or interfere electrically with the edge of the sensor. Shims that are too thick result in an epoxy layer exceeding height specifications, and proper coverage and thermal contact of the epoxy with the sensor cannot be guaranteed. The height of the epoxy layer passed specifications for all interposed modules that were assembled with correct shims. As the large-scale flex PCB lamination processes have stabilized and sites have gained experience with the change in the module assembly procedure, yield losses due to incorrect shimming have decreased.

The primary source of failure for interposed modules was positioning of the power PCB. The power PCB includes a 1-mm-diameter hole in the flex that is used to align the PCB to the tooling used to place the PCB on the module. However, lamination of the flex covers the alignment hole with SE 4445 and kapton, and this hole is no longer accessible. For prototype modules, a makeshift alignment method using shim material as a positional guide for the PCB along its sides was used to align power PCBs to the PCB placement tooling. This method provided only approximate alignment, and any misalignment of the kapton layer with the flex PCB edge can cause an additional shift or rotation in the power PCB placement. If the power PCB is placed too close to the edge of the sensor, electrical issues such as high currents can appear in the sensor. In extreme cases, a shifted or rotated power PCB could prevent proper wirebonding between the readout PCB and the power PCB. 

However, all modules that failed the power PCB positioning specification (8 in the sample of 65 for which data was available) were successfully wirebonded and used for test purposes. New tooling for the power PCB alignment will be fabricated with precisely-positioned pins to ensure more consistent alignment of the power PCB. A combination of adapted tooling and experience gained over time is expected to improve the mechanical quality control yield of interposed modules.

\paragraph{Electrical Quality Control}

Each module undergoes a standard electrical quality control procedure detailed in \cite{ABC130paper,AbeQCProc}. The procedure involves a suite of electrical tests, including a scan to set the pedestal voltage offset for the readout ASIC channels, a scan to set the signal-injection delay for each ASIC, a threshold scan without charge injection to determine the noise occupancy, and a low-bias threshold scan for a fixed injected charge to identify disconnected front-end channels. 

Several tests have emerged as the most indicative of changes in module performance as a result of the interposer design. The ability to bias the sensor is evaluated through the Module IV test, during which the bias voltage is ramped from 0 V to $-$550 V while measuring current through the power PCB. A sensor that passes the module IV test does not experience early HV breakdown (a sharp rise in sensor current) before a bias voltage of $-$500 V. Any noise effects potentially introduced by the new design can be detected through the Three- and Ten-Point-Gain tests, which scan thresholds for three and ten charge injections to obtain gain, noise, and the voltage threshold at 50\% efficiency. These tests are repeated at various temperatures during the module thermal cycling procedure, which includes 10 cycles between $-35$\textdegree C and $+20$\textdegree C in an environmentally-controlled test box. 

Additionally, measurements of the flex PCB NTC temperature readings are used to compare the thermal performance of interposed modules to non-interposed counterparts. Delamination of the interposer layers may also become evident through a rise in flex PCB NTC temperature readings. The flex PCB NTC readings of interposed modules are roughly 1-2\textdegree C higher on readout PCBs and 2-3\textdegree C higher on power PCBs than non-interposed modules. This agrees with a thermal model \cite{Beck:2020qme} that predicts a moderate temperature increase of a few degrees on each PCB. This increase is well within acceptable limits. No temperature rises consistent with delamination have been observed.

\begin{figure}[htbp]
  \centering
  \includegraphics[width=0.8\textwidth]{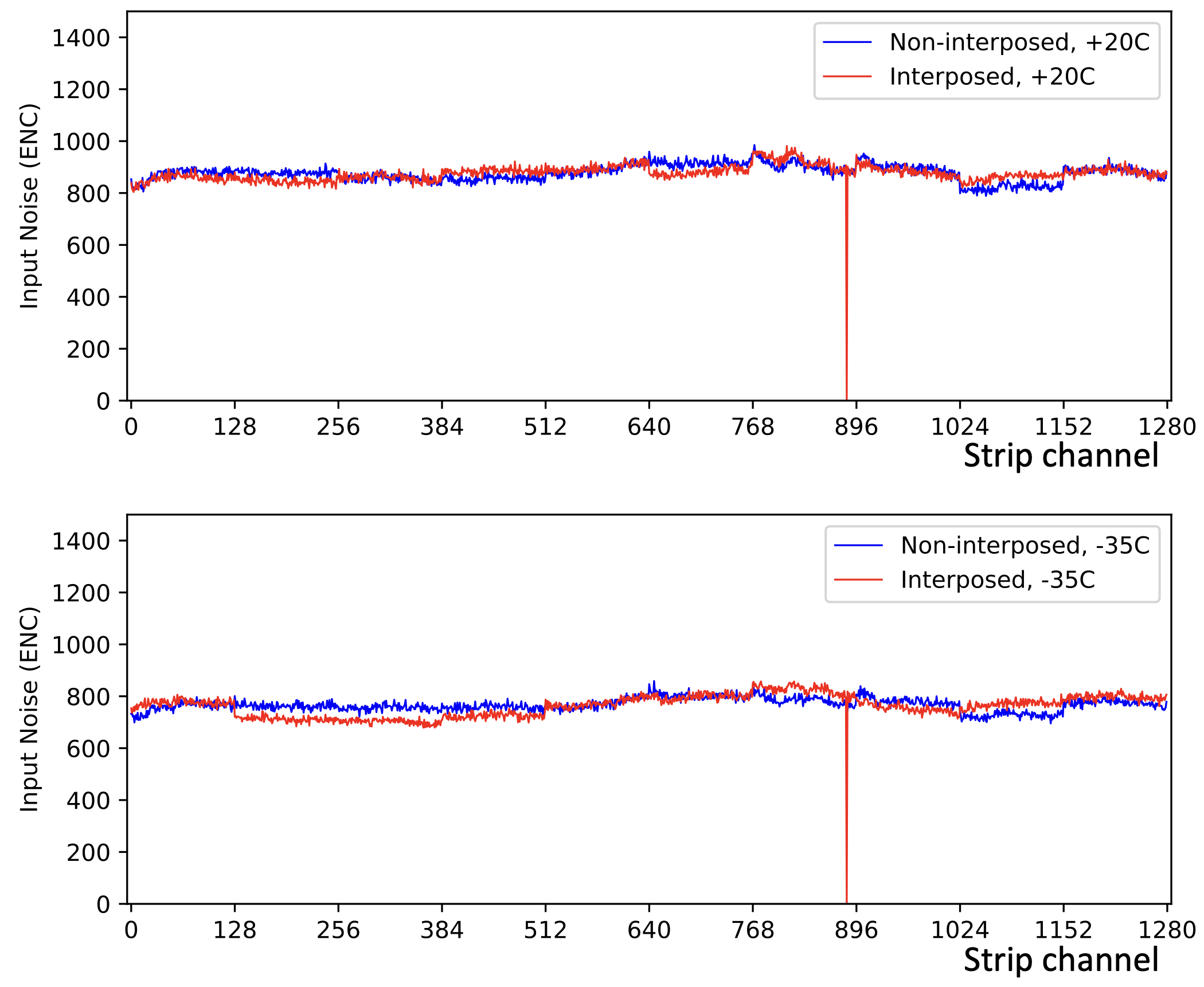}
  \caption{Example comparison of a randomly-selected interposed (red) and non-interposed (blue) long-strip module. The upper plot shows the noise tested at $+20$\textdegree C. The lower plot shows the noise tested at $-35$\textdegree C. No systematic difference in the noise is observed in the interposed module compared to the non-interposed module at either temperature. One channel on a readout ASIC on the interposed module was known to be unresponsive prior to module assembly.}
  \label{fig:noisecomparison}
  \end{figure}

Noise measured through the Three-Point Gain tests was compared between a random selection of interposed and non-interposed modules at room temperature and at $-35$\textdegree C to look for systematic differences. Figure~\ref{fig:noisecomparison} shows a representative example of the noise compared between randomly-selected interposed and non-interposed long-strip modules at $+20$\textdegree C and $-35$\textdegree C. No systematic trend in the noise is observed in the interposed module compared to the non-interposed module. The noise compared between interposed modules can vary by as much as $\pm$ 5\%, which is well within the natural variation observed in non-interposed modules.

During module IV tests, no effect on the overall current draw is evident between interposed and non-interposed modules. Still, similarly to non-interposed modules, early HV breakdown (before $-$500 V) remains a primary failure mode during electrical testing. Dedicated study is ongoing to understand the mechanism behind early HV breakdown. It is hypothesized that some early breakdowns arise from excess epoxy crossing the sensor guard ring, an effect possibly exacerbated by imprecise placement of interposed PCBs. As experience has been gained in assembly of interposed modules, the rate of early breakdown issues has been observed to decrease. Future interposed modules will provide more statistics, and additional experience in interposed module assembly is expected to further reduce yield losses.

\section{Results from Stress Studies}
\label{sec:stresssection}

As mentioned in Section~\ref{sec:solutionsection}, thermo-mechanical simulation predicts an order-of-magnitude decrease in sensor stress with the interposer design. The real-life impact on sensor stress can be determined empirically by comparing the sensor cracking rate of interposed modules loaded on staves with the cracking rate of loaded non-interposed modules, requiring significant statistics. In the meantime, module stress can be quantified through more direct measurements of strain on individual prototype modules. The impact of interposing on sensor stress was studied using thermal deformation of the sensor as a proxy to measure strain.

\subsection{Sensor Deformation after High-Temperature Thermal Cycling}

Module thermal cycling is performed in a box flushed with dry air where modules are placed on temperature-controlled aluminum chucks. Permanent sensor deformation in non-interposed modules has been observed after module thermal cycling if the maximum chuck temperatures exceed $+35^\circ$C~\cite{Salami:2025nob}. This is hypothesized to arise from a glass transition in the epoxy adhesive around $T_\text{glass} \approx 50^\circ$C, which ``bakes in'' sensor deformation caused by CTE mismatch. To prevent this deformation, the nominal module quality control thermal cycling procedure does not include chuck temperatures exceeding $+20^\circ$C. 

A study exploiting this known phenomenon of permanent sensor deformation after high-temperature thermal cycling as a metric of thermal stress is detailed in \cite{Liu:2938335}. Although modules are not expected to experience temperatures above 20\textdegree C when installed in ATLAS, exposing modules to temperatures around $T_\text{glass}$ can be used to explore resilience to thermal deformation. The thermal stress mitigation introduced by the interposer design was hypothesized to reduce measured sensor deformation, which is quantified as sensor \textit{bow}:

\begin{equation}
  |z_\text{bow}| = \text{max}(z_i) - \text{min}(z_i)
  \label{eq:bow-definition}
\end{equation}

where $z$ is measured along the axis normal to the plane of best fit of the sensor.

A control group of five non-interposed long-strip modules was regularly subjected to sensor bow measurements at room temperature between thermal cycling runs. The bow of all five non-interposed modules increased by at least 100~$\mu$m when exposed to $T_\text{max}\geq+35^\circ$C during $[+35, -35]^\circ$C cycling, indicating thermal stress permanently baked into the sensor.

After progressively widening thermal cycling from $[+20, -35]^\circ$C to $12\times [+40, -35]^\circ$C, the bow of all five modules increased by a mean of 146 $\pm$ 27~$\mu$m, but no sensor fractures were observed. However, four out of these five modules exhibited sensor fracturing confirmed by visual inspection after exposing modules to $[+40, -44]^\circ$C cycling. Fractures were observed in two of the modules after the first $[+40, -44]^\circ$C cycle. Sensor fracturing was observed in a third module after an additional set of 10 cycles, and fracturing was observed in the fourth module after a total of 50 $[+40, -44]^\circ$C cycles.

To test the hypothesis of mitigated thermal stress in interposed modules, one long-strip and two short-strip interposed modules were subjected to a similar thermal cycling and bow measurement procedure. No significant change in sensor bow was observed after 10 cycles between $[+40, -44]^\circ$C. One of the short-strip interposed modules was then sent away for test beam studies, while the remaining long-strip and short-strip module underwent 200 cycles with an increasing upper temperature to a maximum temperature range of $[+56, -44]^\circ$C. 
Sensor deformation did appear when $T_\text{max}\geq45^\circ$C, but required significantly more cycles to accumulate than in the non-interposed module case. No signs of mechanical failure in the interposed modules were observed during this study. 

This module thermal cycling study established the conditions required to deform non-interposed modules, with four out of five fracturing after exposure to $[+40, -44]^\circ$C temperatures. Under the same thermal cycling conditions, interposed modules saw little to no sensor deformation. This demonstrates that the stress-decoupling mechanism introduced in the interposer design reduces sensor deformation caused by CTE mismatch. This reduced sensor bow is interpreted as evidence that on average, the sensor in an interposed module is subject to lower thermal stresses.

\subsection{Cold Metrology of Modules Mounted on Glass}

The edge of the sensor is known to curl upward when cold. This curl is related to the strain on the sensor. Measuring the strain experienced by a loaded module is of particular interest. Two interposed modules were loaded onto 6-mm-thick pieces of borosilicate glass using SE 4445 as the loading glue. Borosilicate glass was chosen as a low-CTE structure that would emulate the stave core while allowing visual observation of the loading glue placement. 

\begin{figure}[htbp]
     \centering
     \begin{subfigure}{0.47\textwidth}
         \centering
         \includegraphics[width=\textwidth]{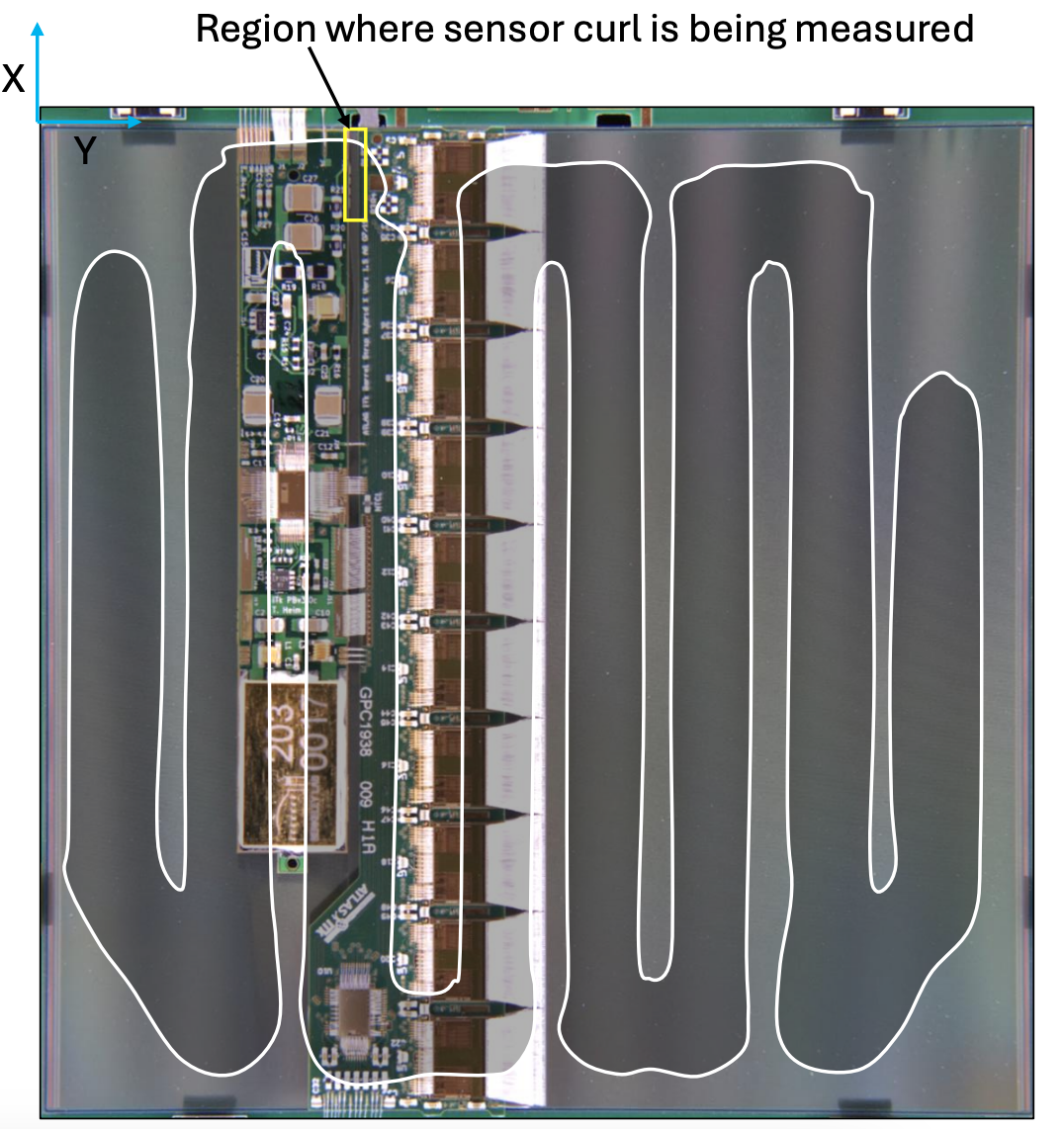}
         \caption{}
         \label{fig:coldmetrology}
     \end{subfigure}
     \begin{subfigure}{0.52\textwidth}
         \centering
         \includegraphics[width=\textwidth]{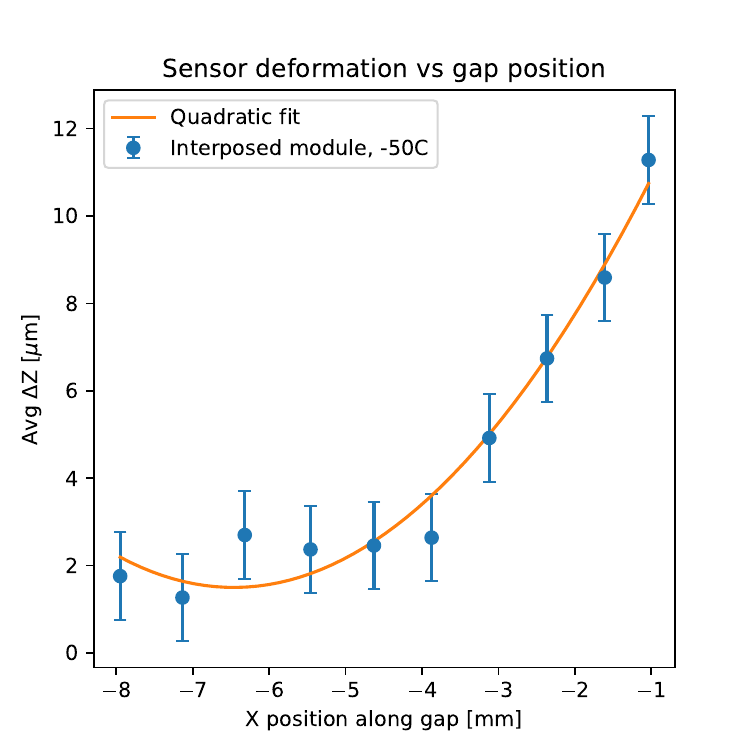}
         \caption{}
         \label{fig:sensorcurl}
     \end{subfigure}
\caption{(a) A module loaded to glass for cold metrology studies. The sensor curl is measured in the indicated yellow boxed region between the PCBs, along the x direction (downward) where x = 0 corresponds to the top edge of the sensor. The loading glue pattern is superimposed on the module to show that the gap between the PCBs is well-supported by the loading glue underneath the module. (b) The vertical displacement of the sensor edge of the long-strip interposed module at $-50$\textdegree C. The curl, or the difference between the maximum and minimum vertical displacement, is $\sim$10 $\mu$m.}
\label{fig:coldmetrologyplots}
\end{figure}

The upward displacement of the sensor is measured along the gap between the readout PCB and the power PCB near the sensor edge, in the region indicated in Figure~\ref{fig:coldmetrology}. The difference between the highest and lowest displacements is taken to be the sensor curl in this region, providing a proxy to strain in the direction parallel to the gap  between the PCBs. 

The displacement measurement of the long-strip module while held at a temperature of $-50$\textdegree C is shown in Figure~\ref{fig:sensorcurl}. At $-50$\textdegree C, the curl of both interposed modules was found to be $\sim$$10$ $\mu$m. In a previous measurement of a glass-loaded non-interposed module at $-50$\textdegree C, the curl was found to be $\sim$$100$ $\mu$m. A significantly smaller curl is observed in the interposed modules, indicating a decrease in strain in loaded interposed modules at cold temperatures compared to non-interposed modules.
 
The curl measurements were repeated after both glass-loaded interposed modules were gamma-irradiated to end-of-life doses. The curl was found to be consistent with the curl of the sensors prior to irradiation within the resolution of the measurement, indicating that the strain on interposed modules does not significantly change after irradiation. This suggests that the stress-relieving effect of the interposer design is retained at end-of-life conditions.

\section{Early Tests of Interposed Modules on Staves}
\label{sec:stavesection}

The ultimate test of the interposer design is whether interposed modules loaded on staves crack when thermally stressed, both for unirradiated and irradiated loaded staves. Prototype interposed stave tests are detailed in \cite{damen2025staveresultsmitigatingsensor}. In an initial test, 70 interposed long-strip modules were loaded onto 2 full staves and one stave side. The staves were thermal cycled between an upper temperature of $+20$\textdegree C and progressively decreasing lower temperatures reaching $-70$\textdegree C. No cracks appeared during thermal cycling within the relevant temperature range of the ITk Strip detector in ATLAS ([$+20$, $-55$]\textdegree C). Below this temperature range, one module exhibited cracking behavior after thermal cycling to $-70$\textdegree C, indicating that the peak sensor stress required for a crack to appear in this population of modules was not reached until well below relevant temperatures. 

\section{Outcome and Conclusion}
\label{sec:outcomesection}

The promising on-stave results opened a path toward restarting module production for the ITk Strip barrel. At the time of writing, the interposer solution is being pursued as the primary cracking mitigation strategy. A series of interposed staves will be assembled at a reduced rate and monitored carefully for issues before ramping up to full production rates.

One remaining question is the electrical performance and cracking rate of loaded interposed modules at end-of-life doses. Irradiation and subsequent thermal cycling of interposed staves are planned to address this.

In the nearly two years since the initial discovery of the cracking problem, the interposer solution has been designed, developed, and tested as a robust stress mitigation strategy. The stress-decoupling design including a layer of kapton and soft glue has been vetted by rigorous material testing, and processes have been developed to implement lamination of flex PCBs at production-ready scales. No significant issues appeared during design validation or quality control testing with prototype interposed modules. Encouraging results from stress studies on individual modules and successful testing of interposed staves have provided a path forward for ITk Strip barrel module production. With the interposer design in place, the ATLAS detector is back on track to have a functioning silicon strip tracker installed for the High-Luminosity LHC.

\acknowledgments

Measurements leading to these results have been performed at the Test Beam Facility at DESY Hamburg (Germany), a member of the Helmholtz Association (HGF). Work performed at SCIPP was supported by the US Department of Energy, grant DE-SC0010107. The work of authors from Lawrence Berkeley National Lab is supported by the Office of High Energy Physics of the U.S. Department of Energy under contract DE-AC02-05CH11231. This work was supported by the Canadian Foundation for Innovation (CFI) and the Natural Sciences and Engineering Research Council of Canada (NSERC).


\bibliographystyle{JHEP}
\bibliography{biblio.bib}

\clearpage
\begin{flushleft}
{\Large The ATLAS ITk Strips Collaboration}

\bigskip

Syed Haider Abidi$^{10}$,
Kirsten Marty Affolder$^{53}$,
Anthony Allen Affolder$^{53}$,
Soren Ahrens$^{16}$,
Torsten Paul Ake \AA kesson$^{35}$,
Theodoros Alexopoulos$^{2}$,
Camron Stephen Alley$^{1}$,
Philip Allport$^{8}$,
Eric Christopher Anderssen$^{6}$,
Egor Antipov$^{57}$
% ...
\bigskip
\\

$^{1}$Department of Physics, University of Adelaide, Adelaide; Australia.\\
$^{2}$Physics Department, National Technical University of Athens, Zografou; Greece.\\
% ...
$^{6}$Physics Division, Lawrence Berkeley National Laboratory, Berkeley CA; United States of America.\\
% ...
$^{8}$School of Physics and Astronomy, University of Birmingham, Birmingham; United Kingdom.\\
% ...
$^{10}$Physics Department, Brookhaven National Laboratory, Upton NY; United States of America.\\
% ...
$^{16}$Deutsches Elektronen-Synchrotron DESY, Hamburg and Zeuthen; Germany.\\
% ...
$^{35}$Fysiska institutionen, Lunds universitet, Lund; Sweden.\\
% ...
$^{53}$Santa Cruz Institute for Particle Physics, University of California Santa Cruz, Santa Cruz CA; United States of America.\\
% ...
$^{57}$Departments of Physics and Astronomy, Stony Brook University, Stony Brook NY; United States of America.\\
% ...

\end{flushleft}

\end{document}